\documentclass{emulateapj}
\usepackage{natbib} 
\slugcomment{Accepted by The Astrophysical Journal}
\shorttitle{DEEP2 Galaxy Clustering}
\shortauthors{Coil et al.}

\begin{document}

\def\mpch {$h^{-1}$ Mpc} 
\def\mpchh {$h^{-1}$ Mpc} 
\def\kms {km s$^{-1}$} 
\def\lcdm {$\Lambda$CDM} 
\def\xisp {$\xi(r_p,\pi)$}
\def\xir{$\xi(r)$}
\def\xis{$\xi(s)$}
\def\rr {$r_0$}
\def\ss {$s_0$}
\def\wprp {$w_p(r_p)$}
\def\sigeight {$\sigma_8^{\rm ~NL}$}

\title{The DEEP2 Galaxy Redshift Survey: Clustering of Galaxies in Early Data}

\author{ 
Alison L. Coil\altaffilmark{1}, 
Marc Davis\altaffilmark{1}, 
Darren S. Madgwick\altaffilmark{1}, 
Jeffrey A. Newman\altaffilmark{1}, 
Christopher J. Conselice\altaffilmark{2}, 
Michael Cooper\altaffilmark{1}, 
Richard S. Ellis\altaffilmark{2}, 
S.~M. Faber\altaffilmark{3}, 
Douglas P. Finkbeiner\altaffilmark{4}, 
Puragra Guhathakurta\altaffilmark{3,6}, 
Nick Kaiser\altaffilmark{5},
David C. Koo\altaffilmark{3}, 
Andrew C. Phillips\altaffilmark{3}, 
Charles C. Steidel\altaffilmark{2}, 
Benjamin J. Weiner\altaffilmark{3},
Christopher N. A. Willmer\altaffilmark{3,7}, 
Renbin Yan\altaffilmark{1}
}

\altaffiltext{1}{Department of Astronomy, University of California,
Berkeley, CA 94720}
\altaffiltext{2}{Department of Astronomy, California Institute of
Technology, Pasadena, CA 91125}
\altaffiltext{3}{University of California Observatories/Lick
Observatory, Department of Astronomy and Astrophysics, University of
California, Santa Cruz, CA 95064}
\altaffiltext{4}{Princeton University Observatory, Princeton, NJ 08544}
\altaffiltext{5}{Institute for Astronomy, University of Hawaii,
Honolulu, HI 96822}
\altaffiltext{6}{Herzberg Institute of Astrophysics, National Research Council of Canada, 5071 West Saanich Road, Victoria, B.C., Canada V9E 2E7}
\altaffiltext{7}{On leave from Observat\'orio Nacional, Rio de Janeiro, Brazil}

\begin{abstract}

We measure the two-point correlation function \xisp \ in a sample of
2219 galaxies between $z=0.7-1.35$ to a magnitude limit of $R_{\rm AB}=24.1$ 
 from the first season of the DEEP2 Galaxy Redshift Survey.  From \xisp \ 
we recover the real-space correlation function, \xir, which we find can be 
approximated within the errors by a power-law,
$\xi(r)=(r/r_0)^{-\gamma}$, on scales $\sim0.1-10$ \mpch.  In a sample
with an effective redshift of $z_{\rm eff}=0.82$, for a \lcdm \ cosmology
we find \rr \ $=3.53 \pm0.81$ \mpch \ (comoving) and $\gamma =1.66
\pm0.12$, while in a higher-redshift sample with $z_{\rm eff}=1.14$ we
find \rr \ $=3.12 \pm0.72$ \mpch \ and $\gamma =1.66 \pm0.12$.  These
errors are estimated from mock galaxy catalogs and are dominated by the
cosmic variance present in the current data sample.  We find that red,
absorption-dominated, passively-evolving galaxies have a larger clustering
scale length, \rr, than blue,
emission-line, actively star-forming galaxies.  Intrinsically brighter 
galaxies also cluster more strongly than fainter galaxies at $z\simeq1$. 
Our results imply that the DEEP2 galaxies have an
effective bias  $b=0.96 \pm0.13$ if $\sigma_{8 \ {\rm DM}}=1$ today or 
$b=1.19 \pm0.16$ if $\sigma_{8 \ {\rm DM}}=0.8$ today.  This bias is
lower than what is predicted by semi-analytic simulations at $z\simeq1$,
which may be the result of our $R$-band target selection.   
We discuss possible 
evolutionary effects within our survey volume, and we compare our
results with galaxy clustering studies at other redshifts, noting that
our star-forming sample at $z\simeq1$ has very similar selection
criteria as the Lyman-break galaxies at $z\simeq3$ and that our red,
absorption-line sample displays a clustering strength comparable to
the expected clustering of the Lyman-break galaxy descendants at $z\simeq1$.
Our results demonstrate that galaxy clustering properties as a function of
color, spectral type and luminosity seen in the local Universe were 
largely in place by $z\simeq1$.

\end{abstract}


\section{Introduction}

Understanding the nature of large-scale structure in the Universe is a key
component of the field of cosmology and is vital to studies of
 galaxy formation and
evolution.  The clustering of galaxies reflects the distribution
of primordial mass fluctuations present in the early Universe and
their evolution with time and also probes the complex physics which
governs the creation of galaxies in their host dark matter potential wells.

Since the first redshift surveys, the two-point correlation function,
\xir, has been used as a measure of the strength of galaxy clustering
\citep{Davis83}.  \xir \ is relatively straightforward to calculate from
pair counts of galaxies, and it has a simple physical interpretation
as the excess probability of finding a galaxy at a separation $r$ from
another randomly-chosen galaxy above that for an unclustered
distribution \citep{Peebles80}.  Locally, \xir \ follows a power-law,
\xir$=(r/r_0)^{-\gamma}$, on scales $\sim1-10$ \mpch \ with $\gamma
\sim1.8$ \citep{Davis83, deLapparent88, Tucker97, Zehavi02,
Hawkins03}. The scale-length of clustering, \rr, is the separation at
which the probability of finding another galaxy is twice the random
probability. Locally \rr \ is measured to be $\sim5.0$ \mpch \  for
optically-selected galaxies but depends strongly on galaxy morphology,
color, type and luminosity \citep{Davis76, Dressler80, Loveday95, 
Hermit96, Willmer98, Norberg01, Zehavi02, Madgwick032df}.

The spatial clustering of galaxies need not trace the underlying
distribution of dark matter.  This was first discussed by
\cite{Kaiser84} in an attempt to reconcile the different clustering
scale lengths of field galaxies and rich clusters, which cannot both
be unbiased tracers of mass.  The galaxy bias, $b$, is a measure of
the clustering in the galaxy population relative to the clustering in
the underlying dark matter distribution.  It can be
defined as the square root of the ratio of the two-point correlation
function of the galaxies relative to the dark matter:
$b=(\xi/\xi_{\rm DM})^{1/2}$, either as a function of $r$ or defined at a
specific scale (see Section 5.1).  
Observations of galaxy clustering have shown that
the galaxy bias can be a function of morphology, type, color,
luminosity, scale and redshift.

Using galaxy morphologies, \cite{Loveday95} find that early-type
galaxies in the Stromlo-APM redshift survey are much more strongly
clustered than late-type galaxies.  Their
early-type sample has a larger correlation length, \rr, and a steeper
slope than late-type galaxies.  However, \cite{Willmer98} show using
data from the Southern Sky Redshift Survey (SSRS2, da Costa et
al. 1998\nocite{daCosta98}) that in the absence of rich clusters 
early-type galaxies have a relative bias of only $\sim1.2$ compared
with late-type galaxies.  In their sample, red galaxies with $(B-R)_0 >
1.3$ are significantly more clustered than blue galaxies, with a
relative bias of $\sim1.4$.  \cite{Zehavi02} also studied galaxy
clustering as a function of color, using data from 
the Sloan Digital Sky Survey (SDSS, York et al. 2000
\nocite{York00}) Early Data Release, and find
that red galaxies ($u^*-r^* > 1.8$) have a larger correlation length, 
\rr, a steeper correlation function, and a larger pairwise velocity
dispersion than blue galaxies. 
They also find a strong dependence of clustering strength on luminosity for 
magnitudes
ranging from $M^*+1.5$ to $M^*-1.5$.  Galaxy clustering for different
spectral types in the 2dF Galaxy Redshift Survey (2dFGRS, Colless et
al. 2001 \nocite{Colless01}) is reported by \cite{Madgwick032df};
 absorption-line galaxies are shown to have a relative bias $\sim2$ times
that of emission-line galaxies on scales $r=1$ \mpch, declining to
unity on larger scales.  Absorption-line galaxies have a steeper
correlation slope and a larger pairwise velocity dispersion.  All of
these results indicate that red, absorption-line, early-type galaxies
are found predominantly in the more massive virialized groups and
clusters in which the random velocities are large.  \cite{Norberg01}
report that the correlation length of optically-selected galaxies in
the 2dFGRS depends weakly on luminosity for galaxies fainter than
$L^*$, the typical luminosity of a galaxy, but rises steeply with
luminosity for brighter galaxies, with the most luminous galaxies
being three times more clustered than $L^*$ galaxies. These results
from local $z\sim0$ surveys indicate that the strength of galaxy
clustering is quite sensitive to different galaxy properties.

 A critical test of both cosmological and galaxy evolution models is
the redshift-dependence of galaxy clustering.  The evolution of the
{\it dark matter} two-point correlation function, $\xi_{\rm DM}(r,t)$,
can be calculated readily and is strongly dependent on cosmology.  In
high-density models the clustering strength grows rapidly, while \lcdm \
models show a more gradual evolution (e.g., Jenkins et al. 1998, Ma 1999)
\nocite{Jenkins98,Ma99}.
However, the evolution of the {\it galaxy} two-point correlation
function, $\xi(r,t)$, depends on the evolution of both the underlying
dark matter distribution and the galaxy bias, which is expected to
increase with redshift.
Applying semi-analytic modelling of galaxy formation and evolution to
dark matter simulations, \cite{Kauffmann99b} present \lcdm \ models with
\rr \ $\sim4$ \mpch \  for the galaxy distribution at $z=1$ compared to
\rr \ $\sim5.2$ \mpch \  locally.  They predict a galaxy bias of $b\sim1.2$
at $z=1$ for galaxies with $M_B<-19+5$ log $(h)$ but also find 
that the galaxy bias can be a strong function of luminosity,
star-formation rate, galaxy type, and sample selection.
\cite{Benson01}, who also apply semi-analytic modelling to \lcdm \ dark
matter simulations, predict a bias of $b=1.5$ at $z=1$ for galaxies
with $M_B < -19.5+5$ log $(h)$. They also predict a similar
morphology-density relation at $z=1$ to that seen locally.

Previous redshift surveys which have attempted to probe intermediate
redshifts from $z=0-1$ have been hampered by small volumes and the
resulting severe cosmic variance.  Results from the Canada-France
Redshift Survey (CFRS, LeFevre et al. 1996) \nocite{LeFevre96} are
based on $\sim600$ galaxies covering 0.14 degrees$^2$, the Norris Redshift 
Survey \citep{Small99} sparsely samples 20 degrees$^2$ with a survey of 
$\sim800$ galaxies, \cite{Hogg00} report on a sample of $\sim1200$ galaxies
 in two very small fields, 
including the Hubble Deep Field, and \cite{Carlberg97} 
present a survey of $\sim250$ galaxies in a total area of 27 arcmin$^2$, 
finding that correlations found in their $K$-band data are generally greater 
than those found by optically-selected surveys. \cite{Small99}
compare results from several surveys which have measured the
correlation length \rr \ in the range $z=0-1$ and illustrate well the
uncertainties in and discrepancies between these results.  For an open
CDM cosmology, the estimates of the comoving correlation length vary
from $\sim2-5$ \mpch \  at $z\simeq0.4-0.6$.  In particular, the
CFRS survey found a much smaller correlation length at $z>0.4$ than the
other surveys, which generally are consistent with weak evolution
between $z=1$ and $z=0$.  A significantly larger survey was undertaken
by CNOC2 \citep{Shepard01}, who obtained redshifts for $\sim5000$\ galaxies 
over 1.44 degrees$^2$.  Most relevant for our purposes may be
 \cite{Adelberger99}, who present clustering results for a deep $R \leq
25.5, z \simeq1$ sample of $\sim800$ galaxies covering a total of 42.5
arcmin$^2$ in 5 fields; they quote a correlation length of \rr \ $\sim3$
\mpch \  for a \lcdm \ cosmology, implying that their galaxy sample is an
unbiased tracer of the mass at $z\simeq1$.  However, many of these surveys
cover very small fields and are likely to underestimate the true
clustering.  There is a well-known systematic bias towards underestimation
of \rr \ in volumes which are small enough that all galaxies are
part of a single large-scale structure and in which the large-scale
modes cannot be sampled.  Furthermore, cosmic 
variance will dominate any measure of
clustering in volumes which are too small to be representative
samples of the Universe \citep{Davis85}.

Here we present early results on galaxy clustering in the DEEP2 Galaxy
Redshift Survey \citep{Davis02}, an $R-$band selected survey which was
designed to study the universe at $z\simeq1$ with a volume and
sampling density comparable to local surveys.  Our intent in this paper 
is to provide an initial measure of the galaxy clustering in our survey at
using the first season of data and to investigate the dependence
of the clustering on galaxy properties, splitting the sample by color,
spectral type, and luminosity.  To constrain galaxy evolution models,
we measure the galaxy bias for the sample as a whole and the relative
bias between subsamples.  This is the first of several planned papers
on galaxy clustering within the DEEP2 survey, and here we focus strictly on
analysis of spatial correlations.  Discussion of redshift-space
distortions will appear in a subsequent paper \citep{Coil04vel}.  In the
data from the first season of observations we measured 5042 redshifts
with $z \ge 0.6$ in three fields with a total area of 0.72
degrees$^2$.  The most complete field currently covers 0.32
degrees$^2$ and includes 2219 galaxies in the redshift range
$z=0.7-1.35$, which we use as the primary data sample in this paper.

The outline of the paper is as follows: in Section 2 we briefly
describe the survey, provide details of the observations, data
reduction, and the data sample used here.  Section 3 outlines the
methods used in this paper, while Section 4 presents our results, both
for the survey sample as a whole and for subsamples based on galaxy
redshift, color, spectral type, and luminosity.  In Section 5 we
discuss galaxy bias and the relative biases between our subsamples,
and we conclude in Section 6.

\section{Data}

\subsection{The DEEP2 Galaxy Redshift Survey}

The DEEP2 Galaxy Redshift Survey is a three-year project using the
DEIMOS spectrograph \citep{Faber02} on the 10-m Keck II telescope to
survey optical galaxies at $z\simeq1$ in a comoving volume of
approximately 6$\times$10$^6$ $h^{-3}$ Mpc$^3$.  The completed survey
will cover 3.5 degrees$^2$ of the sky over four widely separated
fields to limit the impact of cosmic variance.
  The ``1-hour-survey'' (1HS) portion of the DEEP2 project will
use $\sim1$~hr exposure times to 
measure redshifts for $\sim60,000$ galaxies in the redshift range $z
\sim0.7-1.5$ to a limiting magnitude of $R_{\rm AB}=24.1$ (all
magnitudes in this paper are in the AB system; Oke \& Gunn 1983
\nocite{Oke83}).  Photometric data were
taken in $B, R$ and $I$-bands with the CFH12k camera on the 3.6-m
Canada-France-Hawaii telescope.  Galaxies selected for spectroscopy
must additionally meet a color selection given approximately by $B - R
\lesssim 2.35 (R-I) - 0.45$, $R - I \gtrsim 1.15$, or $B - R \lesssim
0.5$.  This simple color-cut was designed to select galaxies at
$z>0.7$ (details are given in Newman et al. 2004\nocite{Newman04}) and
has proven effective in doing so.  As discussed in \cite{Davis02}
this color-cut results in a sample with $\sim$90\% 
of the objects at $z>0.7$, missing only $\sim$5\% of the $z>0.7$ galaxies.

Each of the four DEEP2 1HS fields corresponds to a volume of comoving
dimensions $\sim20 \times 80 \times 
1000$ \mpch \  in a \lcdm \ model at a redshift
of $z=1$. To convert measured redshifts to comoving distances along
the line of sight we assume a flat cosmology with $\Omega_{\rm m}=0.3$
and $\Omega_{\Lambda}=0.7$.  Changing cosmological models within the
range allowed by recent WMAP analysis \citep{Spergel03} has only a
modest influence on our results.  We use $h =  {\rm {\it H}_0/(100 \ km \
s^{-1}})$, and we quote correlation lengths, \rr, in comoving
dimensions of \mpch.

\subsection{Observations and Data Reduction}

This paper uses data from the first observing season of the 1HS portion
of the DEEP2 survey, from August--October 2002.  Three of the four 
DEEP2 fields were observed with a total of 68 custom-made slitmasks. 
Each mask has on the order of $\sim120$ slitlets, with a median 
separation in the spatial direction between targeted galaxies of 
$\sim6$\arcsec, and a minimum of 3\arcsec.  Due to the high source
density of objects, we are able to obtain spectra for $\sim67\%$ of
our targets. Three 20-minute exposures were taken on the
DEIMOS spectrograph with a 1200 line mm$^{-1}$ grating for each
slitmask, covering a spectral range of $\sim6400-9100$ \AA \ at an
effective resolution $R \sim5000$.  The multiple exposures allow us
to robustly reject cosmic rays from the data.  Many of the slitlets in
each mask are tilted to align with the major axis of the target galaxy
to enable internal kinematic studies, and as a result we do not dither
the telescope between exposures.

The data were reduced using a sophisticated IDL pipeline developed at
UC-Berkeley, adapted from spectroscopic reduction
programs developed for the SDSS \citep{Burles04}.  To find the
redshift of each galaxy, a $\chi^2$-minimization is used, where the
code finds minima in $\chi^2$ between the observed spectrum and two
templates; one is an artificial emission-line spectrum convolved with
a broadening function to mimic a 1$\arcsec$ slit and 60 \kms \ internal
dispersion. The other template is a high signal-to-noise ratio
absorption-dominated spectrum which is the average of many thousands
of SDSS galaxies covering a rest wavelength range of 2700-9000 \AA \
\citep{Eisenstein03, Burles04}.  The five most-likely redshifts are saved
and used in a final stage where the galaxy redshift is confirmed by
human interaction.  Our overall redshift success rate is
$\gtrsim$70\% and displays only minor variation with color and 
magnitude ($<20$\%), with the exception of the bluest galaxies 
($R-I<0.4, B-R<0.5$) for which our redshift success rate is $\sim35$\%.
These galaxies represent $\sim25$\% of our targeted sample and account 
for $\sim55$\% of our redshift failures.

The $\lambda$3727 \AA \ [OII] doublet redshifts out of our spectral range at
$z\sim1.44$, and it is believed that all of our bluest ($R-I<0.4, B-R<0.5$) 
targeted galaxies for which we do not measure a redshift lie beyond 
this range.  These galaxies have similar colors and source densities as 
the population at $z\simeq2$ currently studied by C. Steidel and collaborators
(private communication).  If these galaxies were in our observable
redshift window, it is almost certain that we would have measured a redshift,
given that these blue galaxies must have recent star-formation and therefore
 strong emission lines.

Although the instrumental resolution and photon statistics of our data would
suggest that we could achieve a redshift precision of $\sim10$ \kms \ in
the rest frame of each galaxy, we find using galaxies observed twice
on overlapping slitmasks that differences in the position or alignment of 
a galaxy within a slit and internal kinematics within a galaxy lead to an
effective velocity uncertainty of $\sim30$ \kms.

\subsection{Data Sample}

\begin{figure}
\plotone{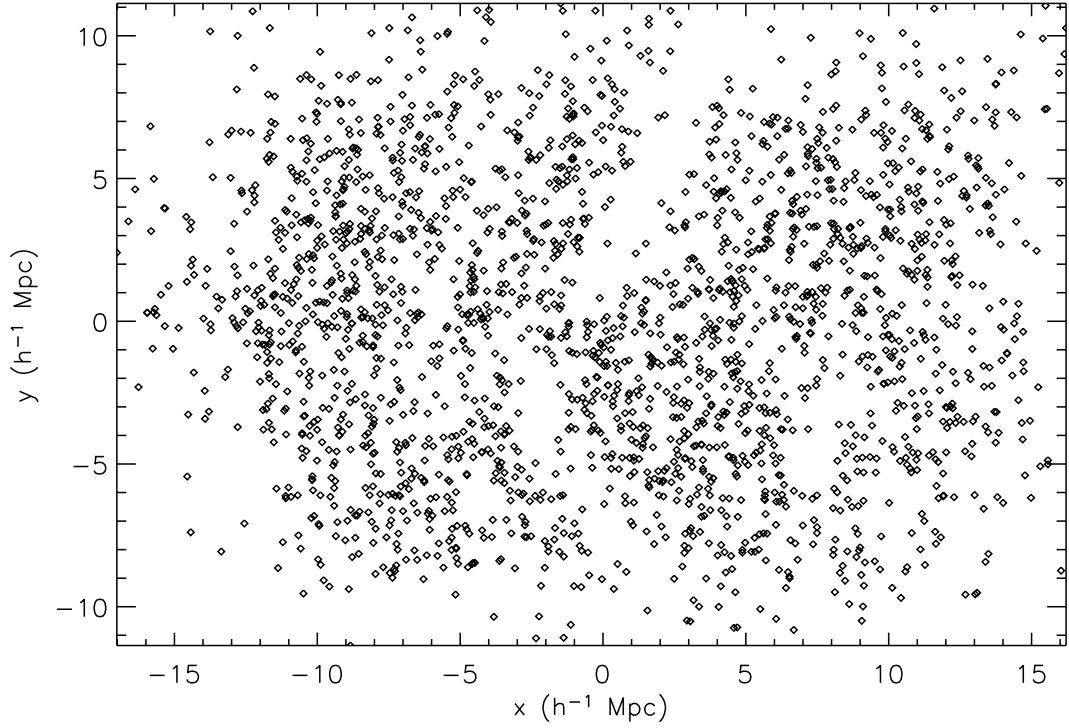}
\caption{Spatial distribution of the full DEEP2 sample of 2219 galaxies 
projected on to the plane of the sky.
\label{spatial}}
\end{figure}

\begin{figure}
\plotone{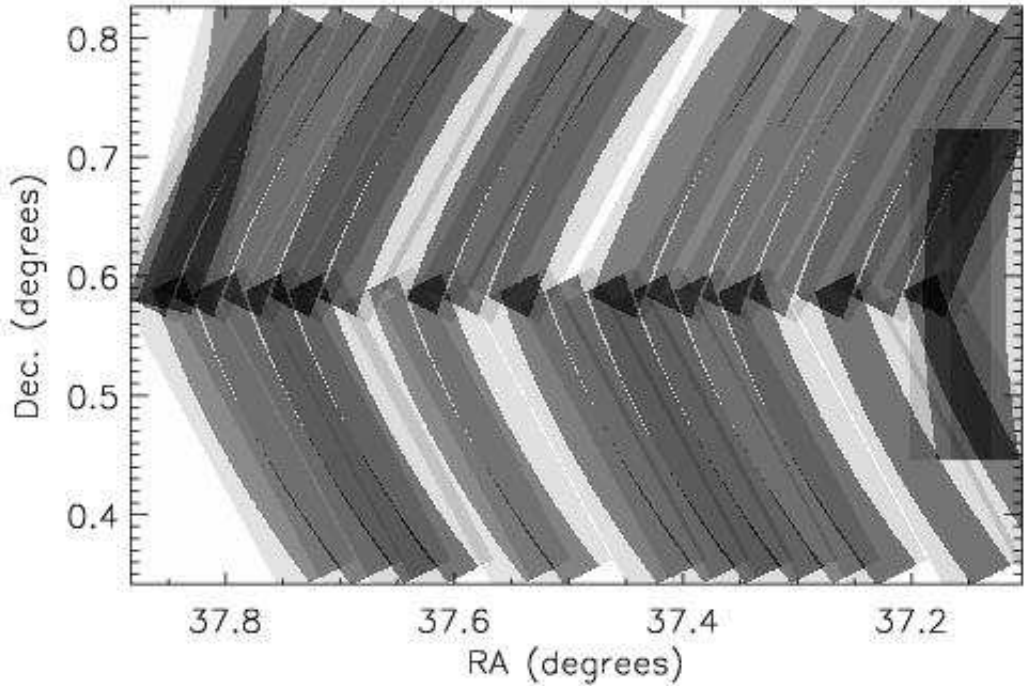}
\caption{Window function of spectroscopic coverage 
in our most complete pointing to date. We include the 32 slitmasks
which have a redshift completeness $\ge 60$\% in our analysis.  The greyscale
ranges from 0 (white) to 0.86 (black) and corresponds to
the probability that a galaxy meeting our selection criteria 
 at that position in the sky was targeted for spectroscopy.  The total
length of this field is 2 degrees; only the first $\sim0.7$ degrees have been
covered thus far.
\label{windowf}}
\end{figure}

Here we present results from only the most complete field, centered at
02 hr 30 min +00 deg, for which we have observed 32 slitmasks
covering $\sim0.7$ degrees by $\sim0.5$ degrees on the sky.
 We use data only from masks which have a redshift success rate of
60\% and higher in order to avoid systematic effects which may bias
our results.  Figures \ref{spatial} and  \ref{windowf} 
show the spatial distribution of galaxies on the plane of the sky and
the window function
for this field.  The observed slitmasks overlap each other in two
horizontal rows on the sky.  Six of the masks have not as yet been
observed in this pointing, leading to regions with lower completeness.

\begin{figure}
\centerline{\scalebox{.85}{\includegraphics{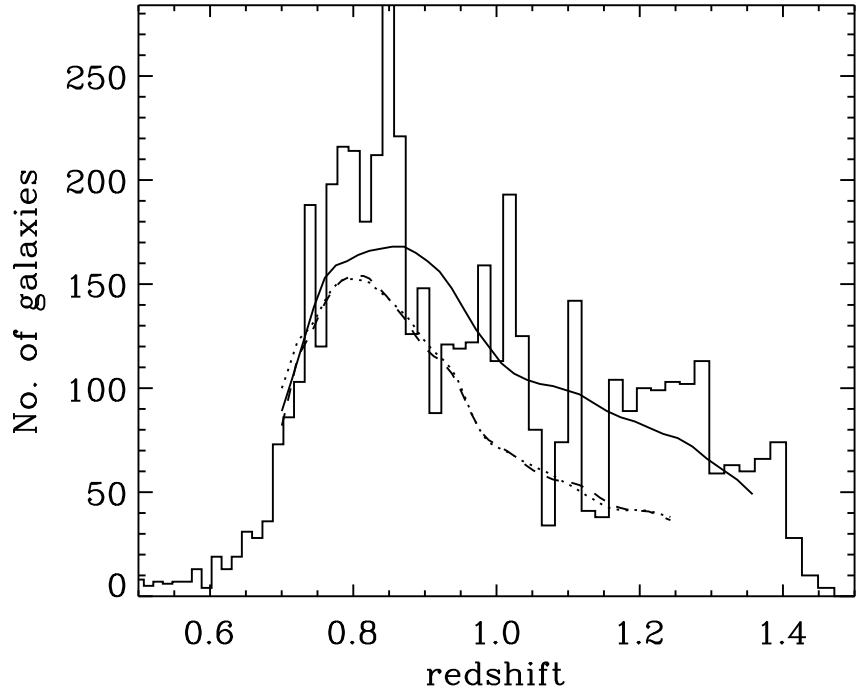}}}
\caption{Redshift distribution of $\sim5000$ galaxies observed in
the first season of the DEEP2 survey, covering three separate fields for
a total of 0.72 degrees$^2$.  
The solid line is a smoothed fit which we use to estimate our 
selection function, $\phi(z)$, in the redshift range $0.7<z<1.35$.
The dotted and dashed lines show the normalized selection functions for the 
emission-line and absorption-line samples, respectively.
\label{zhist}}
\end{figure}

While we measure redshifts as high as $z=1.48$, for this paper we
include only galaxies with $0.7<z<1.35$, a range in which our
selection function is currently well defined.  Our sample in this
field and range contains 2219 galaxies, with a median redshift of
$z=0.90$.  At this median redshift the typical rest-frame wavelength
coverage is $\sim3400-4800$ \AA.  Figure \ref{zhist} shows the overall
redshift distribution of galaxies with $0.5<z<1.5$ in all three of our
observed fields.  There is a rise between redshifts $z=0.7-0.8$, the
result of our probabilistic pre-selection of spectroscopic targets
 expected to have
redshifts $\gtrsim0.7$.  The flux limit of our sample results in the slow
decrease of the observed objects at higher redshifts; smaller-scale
variations are due to galaxy clustering.

In order to compute galaxy correlation statistics, we must understand
our selection function $\phi(z)$, defined as the relative probability
at each redshift that an object will be observed in our sample.  In
general, the selection function can depend on redshift, color,
magnitude, and other properties of the galaxy population and survey
selection.  Ideally one would compute $\phi(z)$ from the luminosity
function of galaxies in the survey.  For this initial study we 
estimate $\phi(z)$ by smoothing the observed redshift
histogram of all the galaxies in our sample, taking into account the 
change in volume with redshift.  We smoothed with a boxcar of 
width 450 \mpch \ and then used an additional boxcar of width 150 \mpch \ 
to ensure that there were no residual bumps due to large-scale structure.  
The resulting $\phi(z)$ is shown by the solid
line in Figure \ref{zhist}.  Also shown in this figure are the normalized 
selection functions for the emission-line and absorption-line samples 
discussed later in the paper (see Section 4.4).  
Note that the redshift distribution $\phi(z)$ is
determined using galaxies in all three of our observed fields, not
only in the field for which we measure \xisp, which reduces effects
due to cosmic variance.  Use of a preliminary $\phi(z)$ constructed
from the luminosity function of our sample does not change the results
presented here.  Using mock catalogs to test the possible systematic
effects due to our estimation of $\phi(z)$, we find that
the resulting error on \rr \ is 5\%, signficantly less than that due to
cosmic variance.

\section{Methods}

\subsection{Measuring the Two-point Correlation Function}

The two-point correlation function \xir \ is a measure of the excess
probability above Poisson of finding a galaxy in a volume element $dV$
at a separation $r$ from another randomly chosen galaxy,
\begin{equation}
dP = n [1+\xi(r)] dV,
\end{equation}
where $n$ is the mean number density of galaxies.  To measure \xir \ one
must first construct a catalog with a random spatial distribution and
uniform density of points with the same selection criteria as the
data, to serve as an unclustered distribution with which to compare
the data.  For each data sample we create a random catalog with
initially $\ge40$ times as many objects with the same overall sky coverage
as the data and uniform redshift coverage.  
This is achieved by applying the window function of our
data, seen in Figure 1, to the random catalog.  Our redshift
completeness is not entirely uniform across the survey; some masks are
observed under better conditions than others and therefore yield a
higher success rate. This spatially-varying redshift success
completeness is taken into account in the window function.  We also
mask the regions of the random catalog where the photometric data had
saturated stars and CCD defects. Finally, we apply our selection
function, $\phi(z)$, so that the random catalog has the same overall
redshift distribution as the data.  This results in a final random
catalog which has $\ge15$ times as many points as the data.  

We measure the two-point correlation function using the
\citet{Landy93} estimator,
\begin{equation}
\xi=\frac{1}{RR}\left[DD \left(\frac{n_R}{n_D}
\right)^2-2DR\left(\frac{n_R}{n_D} \right)+RR\right],
\end{equation}
where $DD, DR$, and $RR$ are pair counts of galaxies in the data-data,
data-random, and random-random catalogs, and $n_D$
and $n_R$ are the mean number densities of galaxies in the data and
random catalogs.  This estimator has been shown to perform as well as the
Hamilton estimator \citep{Hamilton93} but is preferred as it is 
relatively insensitive to the size
of the random catalog and handles edge corrections well
\citep{Kerscher00}.  

As we measure the redshift of each galaxy and not its distance
distortions in $\xi$ are introduced parallel to the line of sight due to
peculiar velocities of galaxies.  On small scales, random motions in
groups and clusters cause an elongation in redshift-space maps along
the line-of-sight known as ``fingers of God''.  On large scales,
coherent infall of galaxies into forming structures causes an apparent
contraction of structure along the line-of-sight \citep{Kaiser87}.
While these distortions can be used to uncover information about the
underlying matter density and thermal motions of the galaxies, they
complicate a measurement of the two-point correlation function in real
space.  Instead, what is measured is \xis, where $s$ is the
redshift-space separation between a pair of galaxies. In order to
determine the effects of these redshift-space distortions and uncover
the real-space clustering properties, we measure $\xi$ in two
dimensions, both perpendicular to and along the line of sight.
Following \cite{Fisher94}, we define ${\bf v_1}$ and ${\bf v_2}$ to be
the redshift positions of a pair of galaxies, ${\bf s}$ to be the
redshift-space separation (${\bf v_1}-{\bf v_2}$), and ${\bf l}
=\frac{1}{2}$(${\bf v_1}+{\bf v_2})$ to be the mean distance to the
pair.  We then define the separation between the two galaxies across
($r_p$) and along ($\pi$) the line of sight as
\begin{equation}
\pi=\frac{{\bf s} \cdot {\bf l}}{{\bf |l|}},
\end{equation}
\begin{equation}
r_p=\sqrt{{\bf s} \cdot {\bf s} - \pi^2}.
\end{equation}
In applying the \citet{Landy93} estimator, we therefore compute pair
counts over a two-dimensional grid of separations to estimate \xisp.

In measuring the galaxy clustering, one sums over counts of galaxy
pairs as a function of separation, normalizing by the counts
of pairs in the random catalog.  While \xisp \  is not a function of the
overall density of galaxies in the sample, if the observed density
is not uniform throughout the sample then a region with higher density 
 will contribute more to the total counts of galaxy pairs,
effectively receiving greater weight in the final calculation.
The magnitude-limit of our survey insures that our selection function,
$\phi(z)$, is not flat, especially at the higher redshift end of our
sample, as seen in Figure 2.  To counteract this, one might weight the
galaxy pairs by $1/\phi(z)$, though this will add significant noise
where $\phi(z)$ is low.  What is generally used instead is the $J_3$
weighting method \citep{Davis82}, which attempts to weight each volume
element equally, regardless of redshift, while minimizing the variance
on large scales.  Using this weighting scheme, each galaxy in a pair 
is given a weight
\begin{equation}
w(z_i,\tau)=\frac{1}{1+4 \pi n_D J_3(\tau) \phi(z_i)},
\label{j3eqn}
\end{equation}
\begin{equation}
J_3(\tau)=\int_{0}^{\tau} \xi(s) s^2 ds,
\end{equation}
where $z_i$ is the redshift of the galaxy, $\tau$ is the
redshift-space separation between the galaxy and its pair object,
$\tau= |s_1-s_2|$, $\phi(z)$ is the selection function of the sample,
such that the mean number density of objects in the sample is $n_D
\phi(z)$ for a homogeneous distribution, and $J_3$ is the volume
integral of \xis.  We limit $\tau \le$ 20 \mpch, the maximum $r_p$
separation we measure, in order to not over-weight the larger scales,
which would lead to a noisier estimate of \xisp.  Note that the
weighting depends on the integral over \xis, a quantity we want to
measure.  Ideally one would iterate the process of estimating \xis \ and
using the measured parameters in the $J_3$ weighting until convergence
was reached.  Here we use a power-law form of
$\xi(s)=(s/s_0)^{-\gamma}$, with initial parameters of \ss \ $=4.4$ \mpch \ 
and $\gamma$=1.5.  These power-law values are in rough accordance with
\xis \ as measured in our full sample.  As tests show that the
measured \xisp \  is quite insensitive to the assumed values of
\ss \ and $\gamma$, we do not iterate this process.  We estimate $n_D$
to be $0.003 h^3$ Mpc$^{-3}$ from the 
observed number density of galaxies in our sample in the redshift
range $z=0.75-0.9$.   As with \ss \ and $\gamma$, we find that the results are
not sensitive to the exact value of $n_D$ used.

\subsection{Deriving the Real-Space Correlations}

While \xis \ can be directly calculated from pair counts, it includes
redshift-space distortions and is not as easily interpreted as \xir,
the real-space correlation function, which measures
only the physical clustering of galaxies, independent of any peculiar
velocities.  To recover \xir \ we use a projection of \xisp \  along
the $r_p$ axis.  As redshift-space distortions affect only the
line-of-sight component of \xisp, integrating over the $\pi$ direction
leads to a statistic \wprp, which is independent of redshift-space
distortions.  Following \cite{Davis83},
\begin{equation}
w_p(r_p)=2 \int_{0}^{\infty} d\pi \ \xi(r_p,\pi)=2 \int_{0}^{\infty}
dy \ \xi(r_p^2+y^2)^{1/2},
\label{eqn}
\end{equation}
where $y$ is the real-space separation along the line of sight. If
\xir \ is modelled as a power-law, $\xi(r)=(r/r_0)^{-\gamma}$, then \rr \ 
and $\gamma$ can be readily extracted from the projected correlation
function, \wprp, using an analytic solution to Equation \ref{eqn}:
\begin{equation}
w_p(r_p)=r_p \left(\frac{r_0}{r_p}\right)^\gamma
\frac{\Gamma(\frac{1}{2})\Gamma(\frac{\gamma-1}{2})}{\Gamma(\frac{\gamma}{2})},
\label{powerlawwprp}
\end{equation}
where $\Gamma$ is the usual gamma function.  A power-law fit to \wprp \
will then recover \rr \ and $\gamma$ for the real-space correlation
function, \xir.  In practice, Equation \ref{eqn} is not integrated to
infinite separations.  Here we integrate to $\pi_{\rm max}=20$ \mpch,
which includes most correlated pairs.  We use analytic
calculations of a broken power-law model for \xir \ which becomes
negative on large scales that we are
underestimating \rr \ by less than $\sim$2\% by not integrating to infinity.

\subsection{Systematic Biases due to Slitmask Observations}

When observing with multi-object slitmasks, the spectra of targets
cannot be allowed to overlap on the CCD array; therefore, objects that
lie near each other in the direction on the sky that maps to the
wavelength direction on the CCD cannot be simultaneously observed.
This will necessarily result in under-sampling the regions with the
highest density of targets on the plane of the sky.  To reduce the
impact of this bias, adjacent slitmasks are positioned approximately a
half-mask width apart, giving each galaxy two chances to appear on a
mask; we also use adaptive tiling of the slitmasks to hold constant
the number of targets per mask. In spite of these steps, the
probability that a target is selected for spectroscopy is diminished
by $\sim25$\% if the distance to its second nearest neighbor is less
than 10 arcseconds (see Davis et al. 2004 for
details).\nocite{Davis04} This introduces a predictable systematic
bias which leads to underestimating the correlation strength on small
scales.

Some previous surveys have attempted to quantify and correct for effects of
this sort using the projected correlation function, $w(\theta)$, of
the sample selected for spectroscopy relative to the that of the entire
photometric sample \citep{Hawkins03}.  Other surveys have attempted to
correct these effects by giving additional weight to observed galaxies
which were close to galaxies which were not observed or by restricting
the scales on which they measure clustering \citep{Zehavi02}.  It is
not feasible for us to use measures of $w(\theta)$, as the
line-of-sight distance that we sample is large ($>$1000 \mpch) and the
resulting angular correlations projected through this distance are
quite small.  In addition, the relation between the decrease in the
2-dimensional angular correlations and the 3-dimensional real-space
correlations is not trivial and depends on both the strength of clustering
and the redshift distribution of sources.

In order to measure this bias, we have chosen to use mock galaxy
catalogs which have similar size, depth, and selection function as our
survey and which simulate the real-space clustering present in our
data.  We have constructed these mock catalogs from the GIF
semi-analytic models of galaxy formation of \cite{Kauffmann99a}. As
described in \cite{Coil01}, we use outputs from several epochs to
create six mock catalogs covering the redshift range $z=0.7-1.5$. To
convert the given comoving distance of each object to a redshift we
assumed a \lcdm \ cosmology; we then constructed a flux-limited
sample which has a similar source density as our data. \cite{Coil01}
presents the selection function and clustering properties of these mock
catalogs.

To quantify the effect of our slitmask target selection on our ability
to measure the clustering of galaxies, we calculate \xisp \ and
\wprp \ in these mock catalogs, both for the full sample of galaxies and
for a subsample that would have been selected to be observed on
slitmasks. The projected correlation function, \wprp, of objects selected
to be on slitmasks is lower on scales $r\le1$ \mpch \ and higher on
scales $r>1$ \mpch \ than the full catalog of objects.  We find
that \rr \ as measured from \wprp \ is overestimated by 1.5\% in the
targeted sample relative to the full sample, while $\gamma$ is
underestimated by 4\%.  Thus our target selection algorithm has a
relatively small effect on estimates of the correlation strength that
is well within the expected uncertainties due to cosmic variance.  
We do not attempt to correct for this effect in this paper.

\section{Results}

\begin{figure}
\centerline{\scalebox{.9}{\rotatebox{90}{\includegraphics{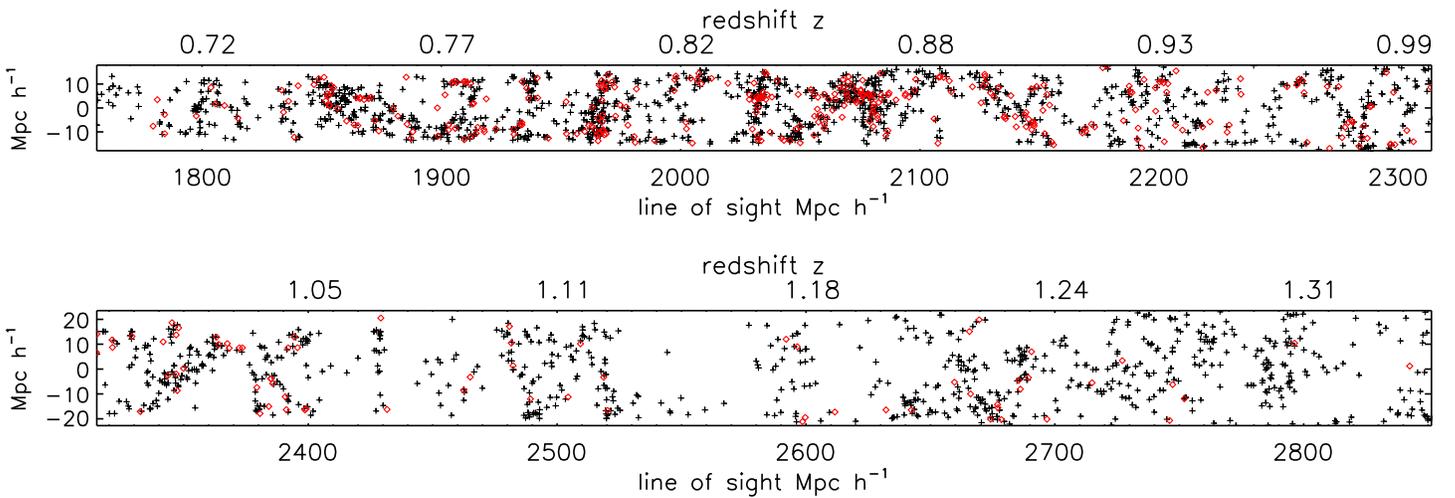}}}}
\caption{Redshift-space distribution of galaxies in early DEEP2 data in
our most complete field shown
as a function of redshift and comoving distance along and projected
distance across the line of sight, assuming a \lcdm \ cosmology. 
We have split the sample by PCA classification: black, plus-signs 
show emission-line galaxies while red, diamond symbols show 
absorption-dominated galaxies.
It is apparent that galaxies with early-type spectra are more strongly
clustered. 
\label{cone}}
\end{figure}

We show the spatial distribution of galaxies in our most complete
field with $0.7<z<1.35$ in Figure \ref{cone}.  We
have projected through the short axis, corresponding to declination,
and plot the comoving positions of the galaxies along and transverse to the
line of sight.  Different symbols show emission-line and
absorption-line galaxies, classified by their spectral type as
discussed in Section 4.4.  Large-scale clustering can be seen, with
coherent structures such as walls and filaments of size $>20$ \mpch \ 
running across our sample.  There are several prominent voids which contain 
very few galaxies, and several overdense regions of strong
clustering.  The visual impression is consistent with \lcdm \ cosmologies
\citep{Kauffmann99b,Benson01}. An analysis of galaxy groups and clusters 
in the early DEEP2 data will be presented by \cite{Gerke04}.

In this paper we focus on measuring the strength of clustering in the galaxy
population using the two-point correlation function.  First we measure
the clustering for the full sample shown in Figure \ref{cone}.  Given
the large depth of the sample in redshift, we then address whether it
is meaningful to find a single measure of the clustering over such an
extended redshift range, as there may be significant evolution in the
clustering strength within our survey volume.  To investigate
evolution within the sample, we would like to measure the clustering
in limited redshift ranges within the survey; given the current sample size, we
divide the data into only two redshift subsamples, studying the front
half and back half of the survey separately.  Finally, we split the full sample
by predicted restframe $(B-R)_0$ color, observed $R-I$ color, spectral
type, and absolute $M_B$ luminosity, to study galaxy clustering as a
function of these properties at $z \simeq1$.  The survey is far from
complete, and with the data presented here we do not attempt to
subdivide the sample further.  In future papers we will be able to
investigate the clustering properties of galaxies in more detail.

\subsection{Clustering in the Full Sample}

\begin{figure}
\centerline{\scalebox{.6}{\includegraphics{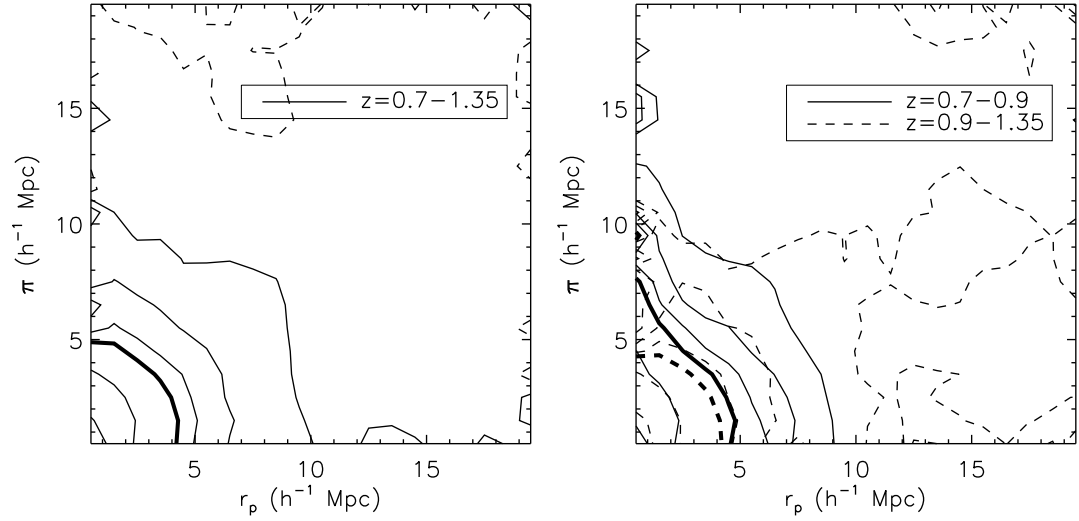}}}
\caption{Left: Contours of the two-dimensional correlation function, \xisp,
smoothed with a $3\times3$ boxcar,  
measured for 2219 galaxies in the redshift range 
$0.7<z<1.35$ in our most complete field to date. The smoothing has
been applied only for the figures; it is not used in calculations. 
Contours levels are 0.0 (dashed), 0.25, 0.5, 0.75, 1.0 (bold), 2.0 and 5.0.
Right: Contours of \xisp, smoothed with a $3\times3$ boxcar, measured for 
lower-redshift galaxies in our sample (solid contours) and for higher-redshift 
objects (dashed contours). Contours levels are 0.25, 0.5, 1.0 (bold), 2.0 and 5.0.
\label{xisp42all}}
\end{figure}

The left side of 
Figure \ref{xisp42all} shows \xisp \  as measured for all galaxies in the
most complete field of our survey in the redshift range $z=0.7-1.35$.
All contour plots presented here have been produced from measurements of 
\xisp \  in linear bins of $1\times1$ \mpch, smoothed with a $3\times3$ boxcar.  
We apply this
smoothing only for the figures; we do not smooth the data before performing
any calculations. On scales $r_p
\le$ 2 \mpch, the signature of small ``fingers of God'' can be seen as
a slight elongation of the contours in the $\pi$ direction.
Specifically, the contours of $\xi=2$ and $\xi=1$ (in bold) intersect
the $\pi$-axis at $\sim3.5$ and 5 \mpch,  while intersecting the
$r_p$-axis at $\sim2.5$ and 4 \mpch, respectively.  We leave a detailed
investigation of redshift-space distortions to a subsequent paper
\citep{Coil04vel}.  

\begin{figure}
\centerline{\scalebox{.6}{\includegraphics{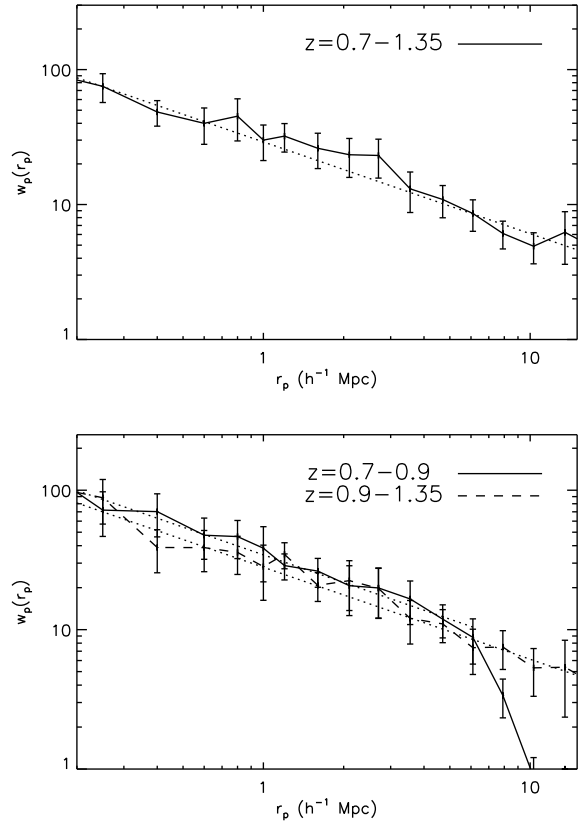}}}
\caption{The projected correlation function, \wprp, for the full
redshift range (top) and two redshift sub-samples (bottom).
The dotted lines show power-law fits used to recover 
\rr \ and $\gamma$ of \xir \ for each sample, as listed in Table 1. 
Error bars are computed from the variance across mock catalogs and
are estimates of the cosmic variance.
\label{wprp42}}
\end{figure}

In order to recover the real-space correlation function, \xir, we 
compute the projected
function \wprp \ by calculating \xisp \  in log-separation bins in $r_p$
and then summing over the $\pi$ direction.  The result is shown in
the top of Figure \ref{wprp42}.  
Errors are calculated from the variance of \wprp \ measured across the six GIF
mock catalogs, after application of this field's current window function. 
Here \wprp \ deviates slightly from a perfect
power-law, showing a small excess on scales $r_p \sim1-3$ \mpch.
However, the deviations are within the 1-$\sigma$ errors, and as there
exists significant covariance between the plotted points, there is no
reason to elaborate the fit.  From \wprp \ we can compute $r_0$ and
$\gamma$ of \xir \ if we assume that \xir \ is a power-law, using
Equation \ref{powerlawwprp}.  Fitting \wprp \ on scales $r_p=0.1-20$
\mpch, we find $r_0=3.19 \pm0.51$ and $\gamma=1.68 \pm0.07$.  This fit
is shown in Figure \ref{wprp42} as the dotted line and is listed in
Table \ref{r0table}.  

The errors on \rr \ and $\gamma$ are taken from the percentage
variance of the measured \rr \ and $\gamma$ amongst the mock catalogs,
scaled to our observed values.  Note that we do not use the errors on
\wprp \ as a function of scale shown in Figure \ref{powerlawwprp},
which have significant covariance, to estimate the errors on \rr \ and
$\gamma$.  The errors quoted on \rr \ and $\gamma$ are dominated by
the cosmic variance present in the current data sample.  To test that
our mock catalogs are independent enough to fully estimate the effects
of cosmic variance, we also measure the variance in new mock catalogs,
which are just being developed \cite{Yan03}.  These new mocks are
made from a simulation with a box size of 300 \mpch \ with finely-spaced
redshift outputs.  We find that the error on \rr \ in the new mocks is
12\%, less than the 16\% we find in the GIF mock catalogs used here.
We therefore believe that the errors quoted here fully reflect the
effects of cosmic variance.  Preliminary measurements in two of our
other fields are consistent with the values found here, within the
1-$\sigma$ errors; in one field, with 1372 galaxies, we measure \rr \
$=3.55$ \mpch \ and $\gamma=1.61$, while a separate field, with 639
galaxies, yields \rr \ $=3.22$ \mpch \ and $\gamma=1.70$.

We have already described above how we use mock catalogs to estimate
the bias resulting from our slitmask target algorithm, which precludes
targeting close pairs in one direction on the sky.  Another method for
quantifying this effect is to calculate an upper-limit on the
clustering using a nearest-neighbor redshift correction, where each
galaxy that was not selected to be observed on a slitmask is given the
redshift of the nearest galaxy on the plane of the sky with a measured
redshift.  This correction will significantly overestimate the
correlations on small scales, since it assumes that members of all 
close pairs on the 
sky are at the same redshift, but it should provide a strong upper
limit on the correlation length \rr.  Using this correction we find an
upper limit on \rr \ of $3.78 \pm0.60$ \mpch \  and on $\gamma$ of 
$1.80 \pm0.07$.

\subsection{Clustering as a Function of Redshift}

In the above analysis, we measured the correlation properties of the full
sample over the redshift range $z=0.7-1.35$.  This is a wide range 
over which to measure a single clustering strength, given
both possible evolutionary effects in the clustering of galaxies
and the changing selection function of our survey, as the luminosity 
distance and the rest-frame bandpass of our selection criteria change with 
redshift.  In addition to
possibly `washing out' evolutionary effects within our survey in
measuring a single clustering strength over this redshift range, the
changing selection function makes it difficult 
to interpret these results.  In this section
we attempt to quantify the redshift dependence in our clustering measurements.

We begin by estimating the effective redshift of the correlation function
 we have measured.  The calculations of \xisp \  presented in the
previous section used the $J_3$ weighting scheme, which attempts to
counteract the selection function of the survey, $\phi(z)$, and give
equal weight to volumes at all redshifts without adding noise.  To
calculate the `effective' redshift $z_{\rm eff}$ of the pair counts used
to calculate \xisp, we compute the mean $J_3$-weighted redshift by
summing over galaxy pairs:
\begin{equation}
z_{\rm eff}= \frac{\sum_{i} \sum_{j,j \not= i} \ z_i \ w_i(z,\tau)^2} {\sum_{i} \sum_{j,j \not= i} \ w_i(z,\tau)^2},
\label{zeff}
\end{equation}
where $i$ runs over all galaxies and $j$ over all galaxies within a range of 
separations from $i$, with $\tau_{min} < \tau < \tau_{max}$, where $\tau$
is the redshift-space separation between the pair of galaxies. The weight
$w_i(z,\tau)$ is given by Equation \ref{j3eqn}, and as both galaxies in 
the pair are at essentially the same redshift (to within $\tau_{max}$ or
better) we use $w_i(z,\tau)^2$ instead of $w_i(z,\tau) w_j(z,\tau)$.
 Note that this effective
redshift will depend on the range of separations considered.  
For $\tau_{min}=1$ \mpch \ and $\tau_{max}=2$ \mpch, 
$z_{\rm eff}=0.96$, while for $\tau_{min}=14$ \mpch \ and $\tau_{max}=16$ 
\mpch, $z_{\rm eff}=1.11$ for the full sample.
For this reason, assuming only one effective redshift for the
correlation function of a deep galaxy sample covering a large redshift
range cannot accurately reflect the true redshift dependence.  We do
however estimate an approximate averaged value for $z_{\rm eff}$
for the galaxy sample presented here by summing
over all pairs of galaxies with $r_p$ or $\pi$ $\le$20 \mpch \ 
($\tau_{min}=0$ \mpch \ and $\tau_{max}=20$ \mpch).  This
yields $z_{\rm eff}=0.99$ for this data sample,  though we
caution that it is not immediately clear how meaningful this number
is, given the wide redshift range of our data.  All values of $z_{\rm eff}$
quoted in Table 1 are for $0 \le \tau \le 20$ \mpch.

If \xisp \  is calculated without $J_3$ weighting, the raw pair counts 
in the survey are dominated by volumes with the highest number 
density in our sample, namely $z=0.75-0.9$.  Without $J_3$ weighting 
we find \rr \ $=3.67 \pm0.59$ \mpch \  and $\gamma=1.65 \pm0.07$.  The 
effective redshift of this result is found using Equation \ref{zeff}, 
setting $w(z)=1$ for all galaxies, yielding $z_{\rm eff}=0.90$.  The 
differences between the values of \rr \ and $\gamma$ derived with and 
without $J_3$ weighting could be the result of evolution within the 
survey sample, cosmic variance, and/or redshift-dependent effects of 
our survey selection.

It is important to stress that our use of the traditional $J_3$
weighting scheme for minimum variance estimates of \xisp \  leads to an
effective redshift of the pair counts which is a function of the pair
separation; the correlations of close pairs have a considerably lower
effective redshift than pairs with large separation.  This complicates
the interpretation of single values of \rr \ and $\gamma$ quoted for the
entire survey.  In local studies of galaxy correlations, one assumes
that evolutionary corrections within the volume studied are
insignificant and that the best correlation estimate will be achieved
with equal weighting of each volume element, provided shot noise does
not dominate.  $J_3$ weighting is intended to provide equal weight per
unit volume to the degree permitted by the radial gradient in source density,
but it complicates interpretation of results within a volume for
which evolutionary effects are expected.  It is far better to
subdivide the sample volume between high and low redshift and
separately apply $J_3$ weighting within the subvolumes.

To this end, we divide our sample near its median redshift, creating
subsamples containing roughly equal numbers of galaxies with
$z=0.7-0.9$ and $z=0.9-1.35$.  The effective redshift for the
lower-$z$ sample (averaged over all separations) is $z_{\rm eff}=0.82$, while
for the higher-$z$ sample it is $z_{\rm eff}=1.14$.  The selection function
for the redshift subsamples is identical to that shown in Figure \ref{zhist},
cut at $z=0.9$.  The right side of
Figure \ref{xisp42all} shows the measured \xisp \ for both subsamples.
At lower redshifts the data exhibit a larger clustering scale length,
as might be expected from gravitational growth of structure.  The
lower-$z$ sample also displays more prominent effects from ``fingers
of God''.  The bottom of Figure \ref{wprp42} shows the resulting \wprp
\ and power-law fits for each redshift range.  Note that we fit a
power-law on scales $r=0.1-20$ \mpch \ for the higher-$z$ sample but
fit on scales $r=0.1-6$ \mpch \ for the lower-$z$ sample, as
\wprp \ decreases significantly on larger scales.  We have tested for
systematic effects which could lead to such a decrease and have not found 
any.  With more data we will be able to see if this dip persists.  
The lower-$z$ sample
exhibits a larger scale length than the higher-$z$
sample, though the difference is well within the 1-$\sigma$
uncertainties due to cosmic variance (see Table \ref{r0table} for details).  
For each subsample we estimate the errors using the variance
among the mock catalogs over the same redshift range used for the data.

A positive luminosity-dependence in the galaxy clustering would lead to an
increase in \rr \ measured at larger redshifts, where the effective
luminosity is greater.  However, evolutionary effects could offset this
effect, if \rr \ grows with time.  We find no significant difference 
in our measured value of \rr \ for the lower-redshift sample. 
Locally, significant luminosity-dependence
has been seen in the clustering of data in the 2dFGRS
\citep{Norberg01} and SDSS \citep{Zehavi02} and, if present in the
galaxy population at $z\simeq1$, could complicate measurements of the
evolution of clustering within our survey volume, given the higher
median luminosity of galaxies in our sample at larger redshifts.  We
investigate the luminosity-dependence of clustering in our sample in
Section 4.5 and discuss possible evolutionary effects in Section 5.2.

\subsection{Dependence of Clustering on Color}

We now measure the dependence of clustering on specific galaxy
properties.  We begin by creating red and blue subsamples based on either
restframe $(B-R)_0$ color or observed $R-I$ color, which is a direct
observable and does not depend on modelling K-corrections.
K-corrections were calculated using a subset of \cite{Kinney96} galaxy
spectra convolved with the $B$, $R$ and $I$ CFH12k filters used in
the DEEP2 Survey. These are used to create a table containing the
restframe colors and K-corrections as a function of $z$ and
$R-I$ color. K-corrections are then obtained for each galaxy using
a parabolic interpolation (for more details see Willmer et al. 2004
\nocite{Willmer04}).  The median K-corrections for the sample used here
are $\sim-0.2$ for $R(z)-B(0)$ and $\sim-0.9$ for $R(z)-R(0)$, where
we apply corrections to our observed $R-$band magnitudes, which are
the deepest and most robustly calibrated \citep{Newman04}.  
After applying these K-corrections, we estimate
the galaxy restframe $(B-R)_0$ colors and divide the sample near the
median color into red, $(B-R)_0 > 0.7$, and blue, $(B-R)_0 < 0.7$, 
subsets.  We further restrict the subsamples to the redshift range
$z=0.7-1.25$.  We fit for the selection function, $\phi(z)$, for each
subsample separately, again using data from all three observed fields
which match these color selection criteria, and find that the 
resulting selection functions for the red and blue subsamples are 
similar to each other and similar to that shown in Figure \ref{zhist}.
For these and all subsamples we use $J_3$ weighting in measuring \xisp
and estimate errors using mock catalogs with half the original sample size.
In this way we attempt to replicate the error due to cosmic variance, 
which depends on the survey volume, and sample size.  We found that the 
error did not increase signficantly when using half the galaxies, which
indicates that cosmic variance is the dominant source of error.

\begin{figure}
\centerline{\scalebox{.8}{\includegraphics{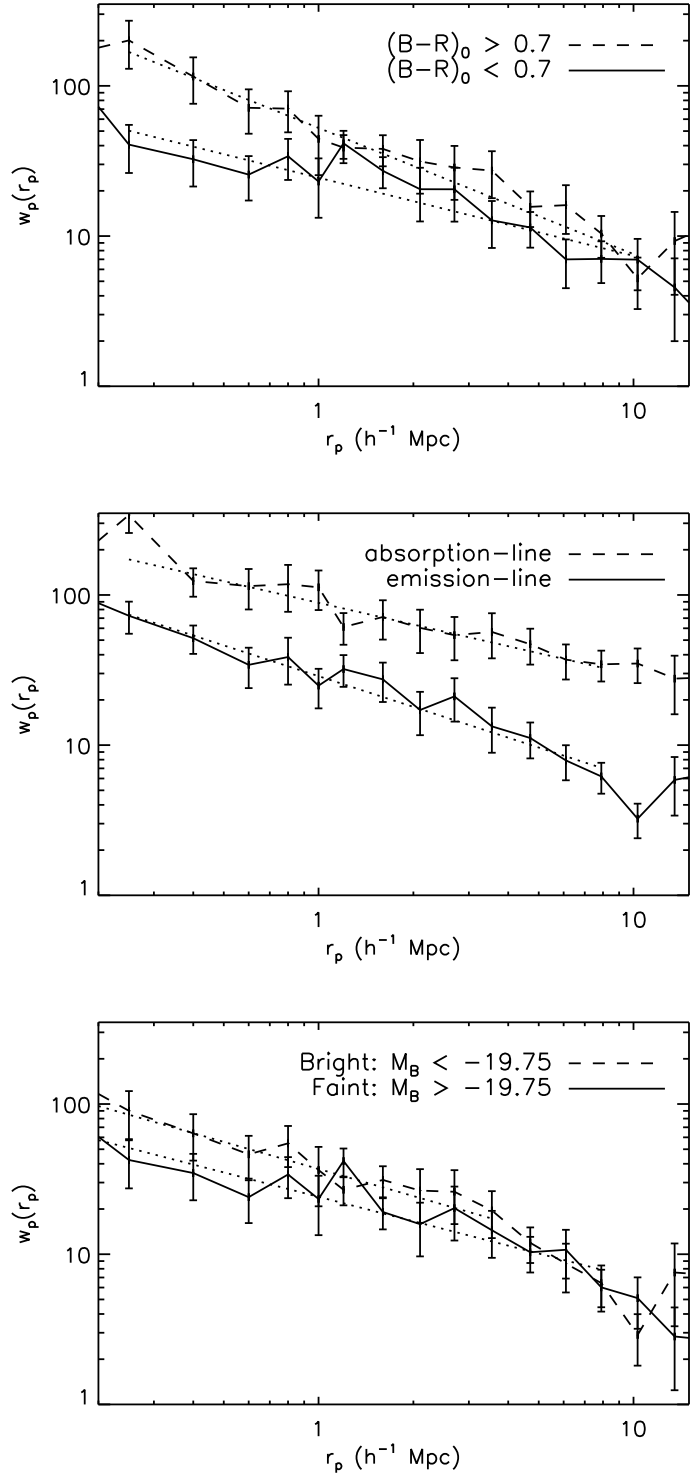}}}
\caption{Top: \wprp \ measured for red (dashed line) and blue (solid line) 
subsamples, divided 
according to restframe $(B-R)_0$ color.  The dotted lines show power-law fits
used to estimate \rr \ and $\gamma$ (see Table 1).  Middle: 
\wprp \ measured for emission-line ($\eta>-13$, solid line) 
and absorption-dominated ($\eta<-13$, dashed line) 
subsamples classified using PCA.
Bottom: \wprp \ measured for subsamples divided according to 
absolute magnitude, $M_B$, assuming $h=1$. 
\label{wprpBR}}
\end{figure}

We find that the red galaxies have a larger correlation length and stronger
``Fingers of God''. This trend is not entirely unexpected, as
previous data at $z\sim1$ have shown similar effects
\citep{Carlberg97, Firth02}, though the volume which we sample is much
larger and therefore less affected by cosmic variance.  The top of
Figure \ref{wprpBR} shows \wprp \ for each sample; fits to \xir \
are given in Table \ref{r0table}.

While restframe colors are more physically meaningful than observed
colors, they are somewhat uncertain as K-corrections can become large
at our highest redshifts.  We therefore divide our full data sample
into red and blue subsets based on observed $R-I$ color. There is a
clear bimodality in the distribution of $R-I$ colors of DEEP2 targets,
leading to a natural separation at $R-I \sim1.1$.
However, as there are not enough galaxies with $R-I> 1.1$ to provide a
robust result, we instead divide the full dataset at $R-I = 0.9$,
which creates subsamples with $\sim4$ times as many blue galaxies as
red.  We again construct redshift selection functions,
$\phi(z)$, for each sample and measure \xisp \  in the redshift range
$z=0.7-1.25$.  Again, the redder galaxies show a larger correlation
length and a steeper slope, though the differences are not as
pronounced as in the restframe color-selected samples; see Table
\ref{r0table}.

\subsection{Dependence of Clustering on Spectral Type}

We next investigate the dependence of clustering on spectral type.
\cite{Madgwick03deep} have performed a principal component analysis
(PCA) of each galaxy spectrum in the DEEP2 survey.  They distinguish
emission-line from absorption-line galaxies using the parameter
$\eta$, the distribution function of which displays a bimodality,
suggesting a natural split in the sample.  We use the same division
employed by \cite{Madgwick03deep}, who define late-type, emission-line
galaxies as having $\eta> -13$ and early-type, absorption-line
galaxies with $\eta< -13$.  Our absorption-line subset includes
$\sim400$ galaxies in the redshift range $z=0.7-1.25$, while the
emission-line sample has $\sim4$ times as many galaxies.  In Figure
\ref{cone} different symbols show the galaxy population divided by
spectral type.  The early-type subset can be seen to reside in the
more strongly clustered regions of the galaxy distribution. 
The middle of Figure \ref{wprpBR} shows \wprp \ measured
for our spectral-type subsamples, and best-fit values of \rr \ and
$\gamma$ are listed in Table \ref{r0table}.  Absorption-line galaxies
have a larger clustering scale length and an increased pairwise
velocity dispersion.  Since $\eta$ correlates well with color, this
result is not unexpected; the bulk of the early-type galaxies have red
colors, though there is a long tail which extends to the median color of 
the late-type galaxies.  Thus the subsamples based on spectral type are 
not identical to those based on color. 
The spectral-type is intimately related to
the amount of current star formation in a galaxy, so that we may
conclude that actively star-forming galaxies at $z\simeq1$ are
significantly less clustered than galaxies that are passively
evolving.  Interestingly, the emission-line galaxies show a 
steeper slope in the correlation function than the absorption-line
galaxies, which is not seen at $z\simeq0$ \citep{Madgwick032df}.  This
will be important to investigate further as the survey collects more
data.

\subsection{Dependence of Clustering on Luminosity}

\begin{figure}
\centerline{\scalebox{.5}{\includegraphics{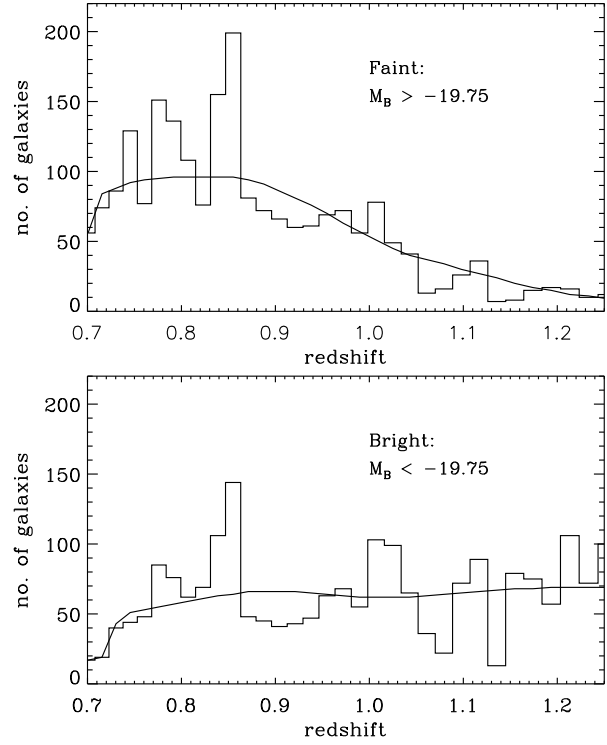}}}
\caption{Redshift histograms and the heavily smoothed curves used to
estimate the selection functions, $\phi(z)$, 
for subsamples divided according to absolute magnitude, $M_B$, assuming $h=1$.  
\label{phiz_lumin}}
\end{figure}

We also split the full sample by absolute $M_B$ magnitude, after
applying K-corrections, to investigate the dependence of galaxy
clustering on luminosity.  We divide our dataset near the median
absolute magnitude, at $M_B=-19.75+$5 log $(h)$.
Figure \ref{phiz_lumin} shows the selection function for each
subsample. Unlike the previous subsets, here the selection functions
are significantly different for each set of galaxies; 
$\phi(z)$ for the brighter objects is relatively flat,
while $\phi(z)$ for fainter galaxies 
falls steeply with $z$.  The bottom of Figure
\ref{wprpBR} shows \wprp \ for each subsample.  We fit \wprp \ as a
power-law on scales $r_p\simeq0.15-4$ \mpch \  and find that the more luminous
galaxies have a larger correlation length.  On larger scales both
samples show a decline in \wprp, but the brighter sample shows a
steeper decline; fits are listed in Table \ref{r0table}.

In this early paper, using a sample of roughly 7\% the size we
expect to have in the completed survey, we have restricted ourselves
to considering only two subsamples at a time.  As a
result, in our luminosity subsamples we are mixing populations of
 red, absorption-line galaxies, which have
very different mass-to-light ratios as well as quite different
selection functions, with the star-forming galaxies that dominate the
population at higher redshifts.  However, the two luminosity
subsamples in our current analysis contain comparable 
ratios of emission-line to 
absorption-line galaxies, with $\sim75$\% of the galaxies in the each
sample having late-type spectra.  In future papers we will be able to
investigate the luminosity-dependence of clustering in
 the star-forming and absorption-line populations separately.

\section{Discussion}

Having measured the clustering strength using the real-space two-point
correlation function, \xir, for each of the samples described above,
we are now in a position to measure the galaxy bias, both for the
sample as a whole at $z\simeq1$ and for subsamples defined by galaxy
properties.  We first calculate the absolute bias for galaxies in our
survey and then determine the relative bias between various subsamples.  Using
these results, we can constrain models of galaxy evolution and compare
our results to other studies at higher and lower redshifts.

\subsection{Galaxy Bias}

To measure the galaxy bias in our sample, we use the parameter
\sigeight, defined as the standard deviation of galaxy count
fluctuations in a sphere of radius 8 \mpch.  We prefer this quantity
as a measure of the clustering amplitude over using the scale-length
of clustering, \rr, alone, which has significant covariance with
$\gamma$.  We can calculate \sigeight \ from a power-law fit to \xir \
using the formula,
\begin{equation}
(\sigma_8^{\rm ~NL})^2 \equiv J_2(\gamma) \left(\frac{r_0}{8 \ h^{-1} {\rm Mpc}}\right)^\gamma,
\label{sig8eqn}
\end{equation}
where
\begin{equation}
J_2(\gamma) = \frac{72}{(3-\gamma)(4-\gamma)(6-\gamma) 2^\gamma}
\end{equation}
 \citep{Peebles80}. Note that here we are not using the linear-theory
$\sigma_8$ that is usually quoted.  Instead, we are evaluating \xir \
on the scale of 8 \mpch \  in the non-linear regime, leading to the
notation \sigeight.

We then define the effective galaxy bias as
\begin{equation}
b = \frac{\sigma_8^{\rm ~NL}}{\sigma_{8 \ {\rm DM}}^{\rm ~NL}},
\end{equation}
where \sigeight \ is for the galaxies and $\sigma_{8 \ {\rm DM}}^{\rm
~NL}$ is for the dark matter.  Our measurements of \sigeight \ for all
data samples considered are listed in Table \ref{r0table}.  Errors are
derived from the standard deviation of \sigeight \ as measured across
the mock catalogs.

The evolution of the dark matter clustering can be predicted readily
using either N-body simulations or analytic theory.  Here we compute
$\sigma_{8 \ {\rm DM}}^{\rm ~NL}$ from the dark matter simulations of
\cite{Yan03} at the effective redshifts of both our lower-$z$ and
higher-$z$ subsamples.  We use two \lcdm \ simulations where the linear
$\sigma_{8 \ {\rm DM}}$ at $z=0$, defined by integrating over the
linear power spectrum, is equal to 1.0 and 0.8. This is the $\sigma_8$
which is usually quoted in linear theory.  For convenience, we define
the parameter $s_8 \equiv \sigma_{8 \ {\rm DM}} (z=0)$.  In both of
these simulations, we fit $\xi(r)_{\rm DM}$ as a power-law on scales
$r \sim 1-8$ \mpch, and from this measure $\sigma_{8 \ {\rm DM}}^{\rm
~NL}$ using equation \ref{sig8eqn} above.  For the simulation with $s_8=1.0$,
 we measure $\sigma_{8 \ {\rm DM}}^{\rm ~NL}=0.70$ at
$z=0.83$ and $\sigma_{8 \ {\rm DM}}^{\rm ~NL}=0.60$ at $z=1.18$, while
for the simulation with $s_8=0.8$, we measure $\sigma_{8 \ {\rm DM}}^{\rm
~NL}=0.56$ at $z=0.83$ and $\sigma_{8 \ {\rm DM}}^{\rm ~NL}=0.49$ at
$z=1.18$.

Our results imply that for $s_8=1.0$ in a \lcdm \ cosmology the 
effective bias of galaxies
in our sample is $b=0.96 \pm0.13$, such that the galaxies trace the
mass.  This would suggest that there was little or no evolution in the
galaxy biasing function from $z=1$ to 0 and could also imply an early
epoch of galaxy formation for these galaxies, such that by $z\simeq1$
they have become relatively unbiased.  However, if $s_8=0.8$ then the
effective galaxy bias in our sample is $b=1.19 \pm0.16$, which is 
more consistent with predictions from semi-analytic models 
\citep{Kauffmann99b}.  Generally, we find the net bias of galaxies in 
our sample to be $b\simeq1/s_8$.

The galaxy bias can be a strong function of sample selection.  One
explanation for the somewhat low clustering amplitude we find may be
the color selection of the survey.  Our flux-limited sample in the
$R$-band translates to bands centered at $\lambda$ = 3600 \AA \ and
3100 \AA \ at redshifts $z=$0.8 and 1.1, respectively.  The flux of a
galaxy at these ultraviolet wavelengths is dominated by young stars,
and therefore our sample could undercount galaxies which have
had no recent star formation, while preferentially selecting galaxies
with recent star formation.
The DEEP2 sample selection may be similar to IRAS-selected low-$z$
galaxy samples in that red, old stellar populations are
under-represented (however, our UV-bright 
sample is probably less dusty than the IRAS galaxies).  
  IRAS-selected samples are known to have a diminished
correlation amplitude and undercount dense regions in cluster cores 
(e.g. Moore et al. 1994)\nocite{Moore94}.  We are accumulating
$K$-band imaging within the DEEP2 fields, which we can use to study
the covariance of $K$-selected samples with our $R$-selected sample in
order to gain a better understanding of the behavior of $\xi(r)$ at $z
\simeq 1$.  \cite{Carlberg97} find that their $K$-selected sample at
$z\sim0.3-1$ generally shows stronger clustering than
optically-selected samples at the same redshifts, and we expect the
same will hold true for our sample.

\subsection{Evolution of Clustering Within Our Survey}

The DEEP2 survey volume is
sufficiently extended in the redshift direction that we expect to
discern evolutionary effects from within our sample.  For example, the
look-back time to $z=0.8$ is 6.9 Gyrs in a \lcdm \ cosmology (for
$h=0.7$), but at $z=1.2$ the look-back time grows to 8.4 Gyrs.  As
discussed in Section 4.2, measuring the clustering strength for the
full sample from $z=0.7-1.35$ is not entirely meaningful, as there may
be significant evolutionary effects within the sample, and the results 
are difficult to
interpret given the dependence of the effective redshift on scale.  We
therefore divide the sample into two redshift ranges and measure the
clustering in the foreground and background of our survey.  However,
with the data available to date, our results must be considered
initial; we hope to report on a sample $\sim20$ times larger in
the next few years.  Note also that as our sample size increases and
we are better able to divide our sample into narrower redshift ranges,
the dependence of $z_{\rm eff}$ on scale will become much less important.

The decreased correlations observed in the higher-redshift subset
within the DEEP2 sample might be considered to be the effect of an
inherent diminished clustering amplitude for galaxies in the more 
distant half of the survey.  Indeed, the mass correlations are expected 
to be weaker at earlier times, but we expect galaxy biasing to be stronger, 
so that the galaxy clustering may not increase with time as the dark matter
distribution does.  As a complication, at
higher redshifts we are sampling intrinsically brighter galaxies due
to the flux limit of the survey, and there is a significant dependence
of clustering strength on luminosity in our data.  Our lower-redshift
sample has an effective luminosity of $M_B=-19.7+$5 log $(h)$, while
for the the higher-redshift galaxies the effective luminosity is 
$M_B=-20.4+$5 log $(h)$.  As discussed in Section 4.2, this luminosity
difference would lead to an increase in \rr \ measured for the higher-redshift
sample, if there was no intrinsic evolution in the galaxy clustering. 
Additionally, at higher redshifts our $R$-band selection corresponds to even shorter
restframe wavelengths, yielding a sample more strongly biased towards
star-forming galaxies.  


\subsection{Comparison with Higher and Lower Redshift Samples}

Galaxies which form at high-redshift are expected to be highly-biased
tracers of the underlying dark matter density field \citep{Bardeen86};
this bias is expected to then decrease with time \citep{Nusser94,
Mo96, Tegmark98}.  If galaxies are born as rare peaks of bias $b_0$ in
a Gaussian noise field with a preserved number density, their bias
will decline with epoch according to $b={(b_0-1)\over D} +1$, where $D$
is the linear growth of density fluctuations in the interval since the
birth of the objects.  This equation shows that if galaxies are highly
biased tracers when born, they should become less biased as the
Universe continues to expand and further structure forms.  This has
been the usual explanation for the surprisingly large clustering
amplitude reported for Lyman-break galaxies at $z\simeq3$.  They have
a clustering scale length comparable to optically-selected galaxies in
the local Universe, but the dark matter should be much less clustered
at that epoch, implying a bias of $b_{\rm Ly B} =4.0 \pm0.7$ for a
\lcdm \ cosmology \citep{Adelberger98}.  The 2dFGRS team has shown that
the bias in their $b_{\rm J}$-selected sample is consistent with
$b_{\rm 2DF} = 1$ \citep{Verde01, Lahav02}.  Given these observations
of $b=4$ at $z\simeq3$ and $b=1$ at $z\simeq0$, one might expect an 
intermediate value of $b$ at $z \simeq 1$, assuming that all of these 
surveys trace similar galaxy populations.  However,
different selection criteria may be necessary to trace the same galaxy
population over various redshifts.

Our subsample of star-forming, emission-line galaxies has similar
selection criteria as recent studies of galaxies at $z\simeq3$.  The
Lyman-break population has been selected to have strong UV
luminosity and therefore high star-formation rates.  The spectroscopic limit of the
Lyman-break sample is $R\sim25.5$, which is roughly equivalent to
$R=23.5$ at $z\sim1$, while the DEEP2 survey limit is $R=24.1$, so
that roughly similar UV luminosities are being probed by these
studies.  With a sample of $\sim700$ Lyman-break galaxies at
$z\simeq3$ \cite{Adelberger03} measure a correlation length of
\rr \ $=3.96 \pm0.29$ \mpch \  with a slope of $\gamma=1.55 \pm0.15$.  At
$z_{\rm eff}=0.99$ we measure a somewhat lower correlation length of
\rr \ $=3.19 \pm0.51$ \mpch \  and a slightly 
steeper slope of $\gamma=1.68 \pm0.07$,
implying that star-forming galaxies at $z\simeq1$ are not as strongly
biased at those at higher redshifts.  The slope of the correlation
function is expected to increase with time, as seen here, as the
underlying dark matter continues to cluster, resulting in more of the
mass being concentrated on smaller scales.  In constraining galaxy
evolution models, however, it is important to note that while these
are measures of similar, star-forming populations of galaxies at
$z\simeq3$ and $z\simeq1$, the Lyman-break galaxies are not
progenitors of the star-forming galaxies at $z\simeq1$.  Using the
linear approximations of \cite{Tegmark98} one would expect the
Lyman-break galaxies to have a correlation length of \rr \ $\sim5$ \mpch \ 
at $z\simeq1$ \citep{Adelberger99}, so that the objects carrying the bulk
of the star formation at $z\simeq1$ and
$z\simeq3$ are not the same.  Our population of red, absorption-line 
galaxies have a
correlation length of \rr \ $\sim5-6$ \mpch, similar to that expected for
the descendants of the Lyman-break population at $z\simeq1$.

Using recent studies from both 2dF and SDSS, we can also compare our
results to $z\simeq0$ surveys.  The two-point correlation function is
relatively well-fit by a power-law in all three of these surveys on
scales $r=1-10$ \mpch.  SDSS find a correlation length of \rr \ $=6.1
\pm0.02$ \mpch \  in their $r^*$-selected sample \citep{Zehavi02}, while
2dF find \rr \ $=5.05 \pm0.26$ \mpch \  in their $b_{\rm J}$-selected
survey.  These values are significantly larger than our measured \rr \ at
$z\simeq1$, in our $R$-selected survey. The slope of the two-point 
correlation function may be marginally steeper at low redshifts, 
with 2dF finding a value of $\gamma=1.67 \pm0.03$ and SDSS
fitting for $\gamma=1.75 \pm0.03$, compared with our values of
$\gamma=1.66 \pm0.12$ at $z_{\rm eff}=0.82$ and $z_{\rm eff}=1.14$.

While the highly-biased, star-forming galaxies seen at $z\simeq3$
appear to have formed in the most massive dark matter halos present at
that epoch \citep{Mo99} and evolved into the red, clustered population
seen at $z\simeq1$, the star-forming galaxies seen at $z\simeq1$ are
not likely to be significantly more clustered in the present Universe.
These galaxies are not highly-biased, and as their clustering properties
 do not imply that they reside in proto-cluster cores, they cannot
become cluster members at $z=0$ in significant numbers.

\subsection{Relative Bias of Subsamples}

Having measured the absolute galaxy bias in our sample as a whole,
which is largely determined by the details of our sample selection, 
we now turn to relative trends seen within our data, which should be
more universal.  Using the various subsamples of our data defined above
we can quantify the dependence of galaxy bias on color, type,
and luminosity, and we compare our findings with other results at
$z=0-1$.

We define the relative bias between two samples as the ratio of their
\sigeight:
\begin{equation}
\frac{b_1}{b_2}\equiv\frac{\sigma_{8 \ 1}^{\rm ~NL}}{\sigma_{8 \
2}^{\rm ~NL}}.
\end{equation}
As the subsamples are taken from the same volume and have similar selection
functions, there is negligable 
cosmic variance in the ratio of the clustering strengths, and
therefore the error in the relative bias is lower than the error on
the values of \sigeight \ individually.  To estimate the error on
the relative bias, we use the variance among the mock catalogs (neglecting
cosmic variance) which leads to a 4\% error, and include an additional error
of 6\% due to uncertainties in the selection function, added in 
quadrature.

We find in the restframe $(B-R)_0$ red and blue subsamples that
$b_{(B-R)_0>0.7}/b_{(B-R)_0<0.7}=1.41 \pm0.10$.  This value is quite
similar to the relative biases seen in local $z=0$ samples.  In the
SSRS2 data \cite{Willmer98} find red galaxies with $(B-R)_0 > 1.3$
have a relative bias of $\sim1.4$ compared to blue galaxies, while
\cite{Zehavi02} report that in the SDSS Early Data Release red
galaxies (based on a split at $u^*-r^*=1.8$) have a relative bias of
$\sim1.6$ compared to blue galaxies.  We find a a similar value of
the relative bias at $z\sim1$ in our red and blue subsamples,
implying that a color-density relation is in place at these higher
redshifts.  The observed-frame $R-I$ subsamples have a relative bias
of $b_{R-I>0.9}/b_{R-I<0.9}=1.29 \pm0.09$.  This value is slightly
lower than that of the restframe color-selected subsamples, as
expected since the observed $R-I$ color of galaxies has a strong
redshift-dependence over the redshift range we cover,
$z=0.7-1.25$, and is therefore less effective at distinguishing
intrinsically different samples.

Using the PCA spectral analysis we find that the absorption-line
sample has a clustering length, \rr, $\sim2$ times larger than the
emission-line sample, with a relative bias of $b_{\rm
absorption-line}/b_{\rm emission-line}=1.77 \pm0.12$.
\cite{Madgwick032df} find using 2dFGRS data that locally,
absorption-line galaxies have a relative bias about twice that of
emission-line galaxies on scales of $r\sim1$ \mpch, but that the
relative bias decreases to unity on scales $>$10 \mpch.  The relative
bias integrated over scales up to 8 \mpch \  is $1.45 \pm0.14$ at $z\simeq0$,
similar to our result at $z\sim1$.  Our current data sample is not
sufficiently large to robustly measure the scale-dependence of the
galaxy bias, though this should readily be measurable from the final
dataset.  \cite{Hogg00} find in their survey (with $z_{\rm med}\sim0.5$)
that galaxies with absorption-line spectra show much stronger
clustering at small separations, though their absorption-line sample
size is small, with 121 galaxies.  \cite{Carlberg97} also report that
in the redshift interval $z=0.3-0.9$, galaxies with red colors have a
correlation length 2.7 times greater than bluer galaxies with strong
[OII] emission.

Recently, there have been several studies which have found very
 large clustering
strengths for extremely-red objects (EROs, $R-K>5$) at $z\sim1$.  Using
the angular correlation function, \cite{Daddi01} find a correlation
length of \rr \ $=12 \pm3$ \mpch \  for EROs at $z\sim1.2$, while
\cite{Firth02} find that 
the correlation length is \rr \ $\sim7.5-10.5$ \mpch.  These
samples are of rare objects which have extreme colors and are quite
luminous; \cite{Firth02} estimate that their sample is $\sim1-1.5$
magnitudes brighter than $M^*$.  We find a correlation length of
\rr \ $=6.61 \pm1.12$ for our absorption-line sample, which has an
effective magnitude of $M_B=-20.5+5$ log$(h)$.  Given
the relatively large clustering strength of the absorption-line
galaxies in our sample and the luminosity difference
between our sample and the ERO studies, it is possible that in
our absorption-line sample we are seeing a somewhat less extreme
population which is related to the EROs seen at $z\sim1$.

We find that the relative bias between luminosity subsamples is
$b_{M_B<-19.75}/b_{M_B>-19.75}=1.24 \pm0.14$.  These datasets have
significantly different selection functions, unlike the previous samples,
and the error on the relative bias due to differences in cosmic variance 
between the two samples results in an additional 8\% error, added in 
quadrature. This is calculated using numerical experiments utilizing
the cosmic variance in redshift bins calculated in \cite{Newman02}.
The bright sample has a
median absolute magnitude of $M_B=-20.3+$5 log $(h)$, while the
faint sample has a median $M_B=-19.1+$5 log $(h)$.  
As noted in Section 4.5, our
luminosity subsamples include both star-forming galaxies as well as
older, absorption-line galaxies and cover a wide range in redshift
($z=0.7-1.25$), possibly complicating interpretation of these
results.  However, both samples have the same ratio of early-type
to late-type spectra.

\section{Conclusions}

The DEEP2 Galaxy Redshift Survey is designed to study the evolution of
the Universe from the epoch $z\sim1.5$ to the present by compiling
an unprecedented dataset with the DEIMOS spectrograph.  With the final
sample we hope to achieve a statistical precision of large-scale
structure studies at $z\sim1$ that is comparable to previous
generations of local surveys such as the Las Campanas Redshift Survey
(LCRS, Shectman et al. 1996\nocite{Shectman96}).  As we complete the
survey, our team will explore the evolution of the properties of
galaxies as well as the evolution of their clustering statistics.

The correlation analysis reported here is far more robust than earlier
studies at $z\sim1$ because of our greatly increased sample size and
survey volume.  We find values of the clustering scale-length,
\rr \ $=3.53 \pm0.81$ \mpch \  at $z_{\rm eff}=0.82$ and \rr \ 
$=3.12 \pm0.72$ \mpch \ 
at $z_{\rm eff}=1.14$, which are within the wide range of clustering
amplitudes reported earlier \citep{Small99, Hogg00}.  This implies a
value of the galaxy bias for our sample, $b=0.96 \pm0.13$ if $\sigma_{8 \ 
{\rm DM}}=1$ today or $b=1.19 \pm0.16$ if $\sigma_{8 \ {\rm DM}}=0.8$ today,
which is lower than what is predicted by semi-analytic simulations 
of $z\simeq1$. Our errors are estimated using mock 
catalogs and are dominated by sample variance, given the current volume 
of our dataset.
 
We find no evidence for significant evolution of \rr \ within our sample,
though intrinsic evolutionary effects could be masked by 
luminosity differences in our redshift subsamples.  
We see a significantly-increased correlation strength for subsets of
galaxies with red colors, early-type spectra, and higher luminosity
relative to the overall population,
similar to the behavior observed in low-redshift catalogs.  Galaxies
with little on-going star formation cluster much more strongly than
actively star-forming galaxies in our sample.  These clustering 
results as a function of color, spectral type and luminosity are
consistent with the trends seen in the semi-analytic simulations of
\cite{Kauffmann99b} at $z=1$, and indicate that galaxy clustering
properties as a function of color, type, and luminosity at $z\sim1$
are generally not very different from what is seen at $z=0$.

The overall amplitude of the galaxy clustering observed within the
DEEP2 survey implies that this is not a strongly biased sample of
galaxies.  For $s_8=1.0$ (defined as the linear $\sigma_{8 \ {\rm
DM}}$ at $z=0$), the galaxy bias is $b=0.96 \pm0.13$, while for
$s_8=0.8$, the bias of the DEEP2 galaxies is $b=1.19 \pm0.16$.  This
low bias may result from the $R$-band selection of the survey,
which roughly corresponds to a restframe $U$-band selected sample; the
more clustered, old galaxies with red stellar populations are likely
to be under-represented as our sample preferentially contains
galaxies with recent star-formation activity.  However, the same selection
bias applies to Lyman-break galaxies studied at $z\simeq3$, which are seen
to be significantly more biased than our sample at $z\simeq1$.

We are undertaking studies with $K$-band data in our fields, which
should lead to clarification of these questions.
 More precise determinations of the
evolution of clustering within our survey and the
luminosity-dependence of the galaxy bias at $z\simeq1$ awaits enlarged
data samples, on which we will report in due course.

\acknowledgements
We would like to thank the anonymous referee for useful comments.
This project was supported in part by the NSF grants AST00-71048, 
AST00-71198 and KDI-9872979. The DEIMOS
spectrograph was funded by a grant from CARA (Keck Observatory), an
NSF Facilities and Infrastructure grant (AST92-2540), the Center for
Particle Astrophysics and by gifts from Sun Microsystems and the Quantum
Corporation. The DEEP2 Redshift Survey has been made possible through
the dedicated efforts of the DEIMOS staff at UC Santa Cruz who built
the instrument and the Keck Observatory staff who have supported it
on the telescope.

\begin{deluxetable}{cccccccc}
\tablecaption{Power-law fits of \xir \ for various data samples. 
\tablenotemark{a}
\label{r0table}} 
\tablehead{ 
\colhead{Sample} & \colhead{no. of} & \colhead{$z$ range} & \colhead{$z_{\rm eff}$}\tablenotemark{b} & 
\colhead{\rr} & 
\colhead{$\gamma$} & \colhead{$r$ range} & \colhead{ $\sigma_8^{\rm NL}$} \\
\colhead{} & \colhead{galaxies} & \colhead{}& \colhead{}&  
 \colhead{\tiny (\mpchh)}& \colhead{}& 
\colhead{\tiny (\mpchh)}& \colhead{} 
}
\startdata

full sample & 2219 & $0.7-1.35$ & 0.99 
   & $3.19 \pm0.51$ & $1.68 \pm0.07$ & $0.1-20$ & $0.60 \pm0.08$ \\


lower $z$ sample & 1087 & $0.7-0.9 $ & 0.82 
  & $3.53 \pm0.81$ & $1.66 \pm0.12$ & $0.1-6$ & $0.66 \pm0.12$\\

higher $z$ sample &1132 & $0.9-1.35$ & 1.14
  & $3.12 \pm0.72$ & $1.66 \pm0.12$ & $0.1-20$ & $0.59 \pm0.11$ \\ 

$(B-R)_0>$ 0.7 & 855 & $0.7-1.25$ & 0.96 
  & $4.32 \pm0.73$ & $1.84 \pm0.07$ & $0.25-10$ & $0.79 \pm0.12$ \\

$(B-R)_0<$ 0.7 & 964 & $0.7-1.25$ & 0.93 
  & $2.81 \pm0.48$ & $1.52 \pm0.06$ & $0.25-10$ & $0.56 \pm0.08$\\

$R-I>$ 0.9 & 442 & $0.7-1.25$ & 0.90
  & $3.97 \pm0.67$ & $1.68 \pm0.07$ & $0.25-8$ & $0.72 \pm0.11$ \\ 
$R-I<$ 0.9 & 1561 & $0.7-1.25$ & 0.95 
  & $2.89 \pm0.49$ & $1.63 \pm0.07$ & $0.25-8$ & $0.56 \pm0.08$ \\

absorption-line & 395 & $0.7-1.25$ & 0.86 
  & $6.61 \pm1.12$ & $1.48 \pm0.06$ & $0.25-8$ & $1.06 \pm0.16$ \\
emission-line & 1605 & $0.7-1.25$ & 0.97 
  & $3.17 \pm0.54$ & $1.68 \pm0.07$ & $0.25-8$ & $0.60 \pm0.09$ \\ 

$M_B < -19.75$ & 899 & $0.7-1.25$ & 0.99 
  & $3.70 \pm0.63$ & $1.60 \pm0.06$ & $0.15-4$ & $0.68 \pm0.10$ \\
$M_B > -19.75$ & 1088 & $0.7-1.25$ & 0.89
  & $2.80 \pm0.48$ & $1.54 \pm0.06$ & $0.15-8$ & $0.55 \pm0.08$ \\

\enddata
\tablenotetext{a}{These fits have not been corrected for the small bias 
we find in our mock catalogs due to our slitmask target selection 
algorithm (see Section 3.3 for details).}
\tablenotetext{b}{see Equation 9}
\end{deluxetable}

\begin{thebibliography}{}

\bibitem[{Adelberger}(1999){Adelberger}]{Adelberger99}
{Adelberger}, K. 1999, In ASP Conference Series Vol. 200

\bibitem[{Adelberger} {et~al.}(1998){Adelberger} {\em et~al.}]{Adelberger98}
{Adelberger}, K.~L., et~al. 1998, \apj, 505, 18

\bibitem[{Adelberger} {et~al.}(2003){Adelberger} {\em et~al.}]{Adelberger03}
{Adelberger}, K.~L., et~al. 2003, \apj, 584, 45

\bibitem[{Bardeen} {et~al.}(1986){Bardeen} {\em et~al.}]{Bardeen86}
{Bardeen}, J.~M., et~al. 1986, \apj, 304, 15

\bibitem[{Benson} {et~al.}(2001){Benson} {\em et~al.}]{Benson01}
{Benson}, A.~J., et~al. 2001, \mnras, 327, 1041

\bibitem[{Burles} \& {Schlegel}(2004){Burles} and {Schlegel}]{Burles04}
{Burles}, S., \& {Schlegel}, D. 2004, in preparation

\bibitem[{Carlberg} {et~al.}(1997){Carlberg} {\em et~al.}]{Carlberg97}
{Carlberg}, R.~G., et~al. 1997, \apj, 484, 538

\bibitem[{Coil} {et~al.}(2004){Coil} {\em et~al.}]{Coil04vel}
{Coil}, A.~L., et~al. 2004, in preparation

\bibitem[{Coil}, {Davis}, \& {Szapudi}(2001){Coil}, {Davis}, and
  {Szapudi}]{Coil01}
{Coil}, A.~L., {Davis}, M., \& {Szapudi}, I. 2001, \pasp, 113, 1312

\bibitem[{Colless} {et~al.}(2001){Colless} {\em et~al.}]{Colless01}
{Colless}, M., et~al. 2001, \mnras, 328, 1039

\bibitem[{da Costa} {et~al.}(1998){da Costa} {\em et~al.}]{daCosta98}
{da Costa}, L.~N., et~al. 1998, \aj, 116, 1

\bibitem[{Daddi} {et~al.}(2001){Daddi} {\em et~al.}]{Daddi01}
{Daddi}, E., et~al. 2001, \aap, 376, 825

\bibitem[{Davis} {et~al.}(2002){Davis} {\em et~al.}]{Davis02}
{Davis}, M., et~al. 2002, Proc. SPIE, 4834, 161 (astro-ph 0209419)

\bibitem[{Davis} {et~al.}(2004){Davis} {\em et~al.}]{Davis04}
{Davis}, M., et~al. 2004, in preparation

\bibitem[{Davis} \& {Geller}(1976){Davis} and {Geller}]{Davis76}
{Davis}, M., \& {Geller}, M.~J. 1976, \apj, 208, 13

\bibitem[{Davis} \& {Huchra}(1982){Davis} and {Huchra}]{Davis82}
{Davis}, M., \& {Huchra}, J. 1982, \apj, 254, 437

\bibitem[{Davis} \& {Peebles}(1983){Davis} and {Peebles}]{Davis83}
{Davis}, M., \& {Peebles}, P.~J.~E. 1983, \apj, 267, 465

\bibitem[{Davis} {et~al.}(1985){Davis}, {Efstathiou}, {Frenk}, and
  {White}]{Davis85}
{Davis}, M., {Efstathiou}, G., {Frenk}, C.~S., \& {White}, S.~D.~M. 1985, \apj,
  292, 371

\bibitem[{de Lapparent}, {Geller}, \& {Huchra}(1988){de Lapparent}, {Geller},
  and {Huchra}]{deLapparent88}
{de Lapparent}, V., {Geller}, M.~J., \& {Huchra}, J.~P. 1988, \apj, 332, 44

\bibitem[{Dressler}(1980){Dressler}]{Dressler80}
{Dressler}, A. 1980, \apj, 236, 351

\bibitem[{Eisenstein} {et~al.}(2003){Eisenstein} {\em et~al.}]{Eisenstein03}
{Eisenstein}, D.~J., et~al. 2003, \apj, 585, 694

\bibitem[{Faber} {et~al.}(2002){Faber} {\em et~al.}]{Faber02}
{Faber}, S., et~al. 2002, Proc. SPIE, 4841, 1657

\bibitem[{Firth} {et~al.}(2002){Firth} {\em et~al.}]{Firth02}
{Firth}, A.~E., et~al. 2002, \mnras, 332, 617

\bibitem[{Fisher} {et~al.}(1994){Fisher} {\em et~al.}]{Fisher94}
{Fisher}, K.~B., et~al. 1994, \mnras, 267, 927

\bibitem[{Gerke} {et~al.}(2004){Gerke} {\em et~al.}]{Gerke04}
{Gerke}, B., et~al. 2004, in preparation

\bibitem[{Hamilton}(1993){Hamilton}]{Hamilton93}
{Hamilton}, A.~J.~S. 1993, \apj, 417, 19

\bibitem[{Hawkins} {et~al.}(2003){Hawkins} {\em et~al.}]{Hawkins03}
{Hawkins}, E., et~al. 2003, \mnras, 346, 78

\bibitem[{Hermit} {et~al.}(1996){Hermit} {\em et~al.}]{Hermit96}
{Hermit}, S., et~al. 1996, \mnras, 283, 709

\bibitem[{Hogg}, {Cohen}, \& {Blandford}(2000){Hogg}, {Cohen}, and
  {Blandford}]{Hogg00}
{Hogg}, D.~W., {Cohen}, J.~G., \& {Blandford}, R. 2000, \apj, 545, 32

\bibitem[{Jenkins} {et~al.}(1998){Jenkins} {\em et~al.}]{Jenkins98}
{Jenkins}, A., et~al. 1998, \apj, 499, 20

\bibitem[{Kaiser}(1984){Kaiser}]{Kaiser84}
{Kaiser}, N. 1984, \apjl, 284, L9

\bibitem[{Kaiser}(1987){Kaiser}]{Kaiser87}
{Kaiser}, N. 1987, \mnras, 227, 1

\bibitem[{Kauffmann} {et~al.}(1999a){Kauffmann}, {Colberg}, {Diaferio}, and
  {White}]{Kauffmann99a}
{Kauffmann}, G., {Colberg}, J.~M., {Diaferio}, A., \& {White}, S.~D.~M. 1999a,
  \mnras, 303, 188

\bibitem[{Kauffmann} {et~al.}(1999b){Kauffmann}, {Colberg}, {Diaferio}, and
  {White}]{Kauffmann99b}
{Kauffmann}, G., {Colberg}, J.~M., {Diaferio}, A., \& {White}, S.~D.~M. 1999b,
  \mnras, 307, 529

\bibitem[{Kerscher}, {Szapudi}, \& {Szalay}(2000){Kerscher}, {Szapudi}, and
  {Szalay}]{Kerscher00}
{Kerscher}, M., {Szapudi}, I., \& {Szalay}, A.~S. 2000, \apjl, 535, L13

\bibitem[{Kinney} {et~al.}(1996){Kinney} {\em et~al.}]{Kinney96}
{Kinney}, A.~L., et~al. 1996, \apj, 467, 38

\bibitem[{Lahav} {et~al.}(2002){Lahav} {\em et~al.}]{Lahav02}
{Lahav}, O., et~al. 2002, \mnras, 333, 961

\bibitem[{Landy} \& {Szalay}(1993){Landy} and {Szalay}]{Landy93}
{Landy}, S.~D., \& {Szalay}, A.~S. 1993, \apj, 412, 64

\bibitem[{Le Fevre} {et~al.}(1996){Le Fevre} {\em et~al.}]{LeFevre96}
{Le Fevre}, O., et~al. 1996, \apj, 461, 534

\bibitem[{Loveday} {et~al.}(1995){Loveday}, {Maddox}, {Efstathiou}, and
  {Peterson}]{Loveday95}
{Loveday}, J., {Maddox}, S.~J., {Efstathiou}, G., \& {Peterson}, B.~A. 1995,
  \apj, 442, 457

\bibitem[{Ma}(1999){Ma}]{Ma99}
{Ma}, C. 1999, \apj, 510, 32

\bibitem[{Madgwick} {et~al.}(2003a){Madgwick} {\em et~al.}]{Madgwick032df}
{Madgwick}, D.~S., et~al. 2003a, \mnras, 344, 847

\bibitem[{Madgwick} {et~al.}(2003b){Madgwick} {\em et~al.}]{Madgwick03deep}
{Madgwick}, D.~S., et~al. 2003b, \apj, 599, 997

\bibitem[{Mo} \& {White}(1996){Mo} and {White}]{Mo96}
{Mo}, H.~J., \& {White}, S.~D.~M. 1996, \mnras, 282, 347

\bibitem[{Mo}, {Mao}, \& {White}(1999){Mo}, {Mao}, and {White}]{Mo99}
{Mo}, H.~J., {Mao}, S., \& {White}, S.~D.~M. 1999, \mnras, 304, 175

\bibitem[{Moore} {et~al.}(1994){Moore} {\em et~al.}]{Moore94}
{Moore}, B., et~al. 1994, \mnras, 269, 742

\bibitem[{Newman} {et~al.}(2004){Newman} {\em et~al.}]{Newman04}
{Newman}, J., et~al. 2004, in preparation

\bibitem[{Newman} \& {Davis}(2002){Newman} and {Davis}]{Newman02}
{Newman}, J.~A., \& {Davis}, M. 2002, \apj, 564, 567

\bibitem[{Norberg} {et~al.}(2001){Norberg} {\em et~al.}]{Norberg01}
{Norberg}, P., et~al. 2001, \mnras, 328, 64

\bibitem[{Nusser} \& {Davis}(1994){Nusser} and {Davis}]{Nusser94}
{Nusser}, A., \& {Davis}, M. 1994, \apjl, 421, L1

\bibitem[{Oke} \& {Gunn}(1983){Oke} and {Gunn}]{Oke83}
{Oke}, J.~B., \& {Gunn}, J.~E. 1983, \apj, 266, 713

\bibitem[{Peebles}(1980){Peebles}]{Peebles80}
{Peebles}, P.~J.~E. 1980.
\newblock {The Large-Scale Structure of the Universe}, (Princeton, N.J.,
  Princeton Univ. Press)

\bibitem[{Shectman} {et~al.}(1996){Shectman} {\em et~al.}]{Shectman96}
{Shectman}, S.~A., et~al. 1996, \apj, 470, 172

\bibitem[{Shepherd} {et~al.}(2001){Shepherd} {\em et~al.}]{Shepard01}
{Shepherd}, C.~W., et~al. 2001, \apj, 560, 72

\bibitem[{Small} {et~al.}(1999){Small}, {Ma}, {Sargent}, and
  {Hamilton}]{Small99}
{Small}, T.~A., {Ma}, C., {Sargent}, W.~L.~W., \& {Hamilton}, D. 1999, \apj,
  524, 31

\bibitem[{Spergel} {et~al.}(2003){Spergel} {\em et~al.}]{Spergel03}
{Spergel}, D.~N., et~al. 2003, \apjs, 148, 175

\bibitem[{Tegmark} \& {Peebles}(1998){Tegmark} and {Peebles}]{Tegmark98}
{Tegmark}, M., \& {Peebles}, P.~J.~E. 1998, \apjl, 500, L79

\bibitem[{Tucker} {et~al.}(1997){Tucker} {\em et~al.}]{Tucker97}
{Tucker}, D.~L., et~al. 1997, \mnras, 285, L5

\bibitem[{Verde} {et~al.}(2002){Verde} {\em et~al.}]{Verde01}
{Verde}, L., et~al. 2002, \mnras, 335, 432

\bibitem[{Willmer} {et~al.}(2004){Willmer} {\em et~al.}]{Willmer04}
{Willmer}, C., et~al. 2004, in preparation

\bibitem[{Willmer}, {da Costa}, \& {Pellegrini}(1998){Willmer}, {da Costa}, and
  {Pellegrini}]{Willmer98}
{Willmer}, C.~N.~A., {da Costa}, L.~N., \& {Pellegrini}, P.~S. 1998, \aj, 115,
  869

\bibitem[{Yan} {et~al.}(2003){Yan} {\em et~al.}]{Yan03}
{Yan}, R., et~al. 2003, accepted by ApJ (astro-ph/0311230)

\bibitem[{York} {et~al.}(2000){York} {\em et~al.}]{York00}
{York}, D.~G., et~al. 2000, \aj, 120, 1579

\bibitem[{Zehavi} {et~al.}(2002){Zehavi} {\em et~al.}]{Zehavi02}
{Zehavi}, I., et~al. 2002, \apj, 571, 172

\end{thebibliography}
\end{document}